\documentclass[11pt,3p]{elsarticle}
\usepackage{verbatim}
\usepackage{lineno,hyperref}
\modulolinenumbers[5]
\usepackage{color}
\journal{CARBON (Elsevier)}
\usepackage{siunitx}
\usepackage[]{subfigure}
\usepackage{placeins} 
\usepackage[english]{babel}  
\usepackage[utf8x]{inputenc}
\usepackage{graphicx}
\usepackage{booktabs}
\usepackage{tabularx}
\usepackage{caption}
\usepackage{amsmath}
\usepackage{mathtools}
\usepackage{float}

\biboptions{authoryear}

\begin{document}

\begin{frontmatter}

\title{Designing graphene based nanofoams with nonlinear auxetic and anisotropic mechanical properties under tension or compression}

\author[DICAM,TIPFA]{Andrea Pedrielli}

\author[TIPFA,PRA]{Simone Taioli}

\author[TIPFA]{Giovanni Garberoglio\corref{mycorrespondingauthor2}}
\cortext[mycorrespondingauthor2]{Corresponding author}
\ead{garberoglio@ectstar.eu}

\author[DICAM,LONDON,BIO]{Nicola Pugno\corref{mycorrespondingauthor1}}
\cortext[mycorrespondingauthor1]{Corresponding author}
\ead{nicola.pugno@unitn.it}

\address[DICAM]{Laboratory of Bio-inspired \& Graphene Nanomechanics, Department of Civil, Environmental and Mechanical Engineering, University of Trento, Via Mesiano 77, 38123 Trento, Italy}
\address[TIPFA]{European Center for Theoretical Studies in Nuclear Physics and Related Areas (ECT*-FBK) and Trento Institute for Fundamental Physics and Applications (TIFPA-INFN), 38123 Trento, Italy}
\address[PRA]{Faculty of Mathematics and Physics, Charles University in Prague, 180 00 Prague 8, Czech Republic}
\address[LONDON]{School of Engineering and Materials Science, Materials Research Institute, Queen Mary University of London, London E1 4NS, UK}
\address[BIO]{Center for Materials and Microsystems, Fondazione Bruno Kessler, Via Sommarive 18, 38123 Povo, TN, Italy}

\begin{abstract}
In this study, the analysis of the mechanical response of realistic
fullerene-nanotube nanotruss networks with face-centered cubic geometry is
performed by using molecular dynamics with reactive potentials. In
particular, the mechanical properties of these novel architectures are
investigated in both compressive and tensile regimes, by straining along
different directions a number of truss geometries. Our atomistic
simulations reveal a similar behavior under tensile stress for all the
samples. Conversely, under compressive regimes the emergence of a response
that depends on the orientation of load is observed together with a peculiar
local instability. Due to this instability, some of these nanotruss
networks present a negative Poisson ratio in compression, like re-entrant
foams. Finally, the performance of these nanotruss networks is analyzed
with regards to their use as impact energy absorbers, finding properties
outperforming materials traditionally used in these applications.
\end{abstract}

\end{frontmatter}
\section{Introduction}

Recent advances in single and multi-layered graphene growth techniques
\citep{Avouris2012, Tatti2016} have renewed the interest in synthesising
carbon-based porous nanomaterials. These materials are promising for a
broad range of applications, ranging from energy storage in amorphous
structures \citep{Alonso2012} to tunable hierarchical nanotube scaffolds
for regenerative medicine \citep{Coluci2010}, and lightweight foams for oil
absorption \citep{Wang2014}. Furthermore, pristine graphene shows exceptional mechanical properties, such as the the biggest ultimate strength ever found and the ability to retain its initial size after strain \citep{Frank2007}. However, transferring these unique properties to macroscale still represents a challenge for materials scientists. For reliable structural applications it is essential to build macroscopic 3D architectures
preserving the intrinsic properties of the material. This can be achieved
through a proper tuning of the porosity and cell geometry which are the parameters that mainly affect the mechanical properties of porous materials \citep{Fleck2010}.

Carbon nanomaterials with random porosity distribution were initially
proposed as possible means to transfer graphene's unique mechanical
properties, such as Young modulus, tensile strength and toughness from
nano- to macro-scale. Unfortunately, it turned out that random-pore
structures actually exhibit poor scaling of these mechanical properties
with decreasing density \citep{Hodge2005}. Furthermore, they are limited in
number of achievable architectures, due to the uncontrollable porous
distribution.

At variance, in periodic architectures one can expect all nanoscopic
components to work in synergy, and thus deliver optimal mechanical
properties at larger length scales.  In particular, ordered 3D
nano-architectures can be designed to realize specific functional
properties such as negative \citep{Hall2008} or flipping Poisson ratio
\citep{Wu2013} or to obtain a significant increase in gas and energy
storage \citep{Ding2007,Tylianakis2011, Garberoglio2015,taiol}.

\begin{figure*}[]
\centering
\subfigure[FCC unit cell]
{\includegraphics[width=%
0.25\textwidth]{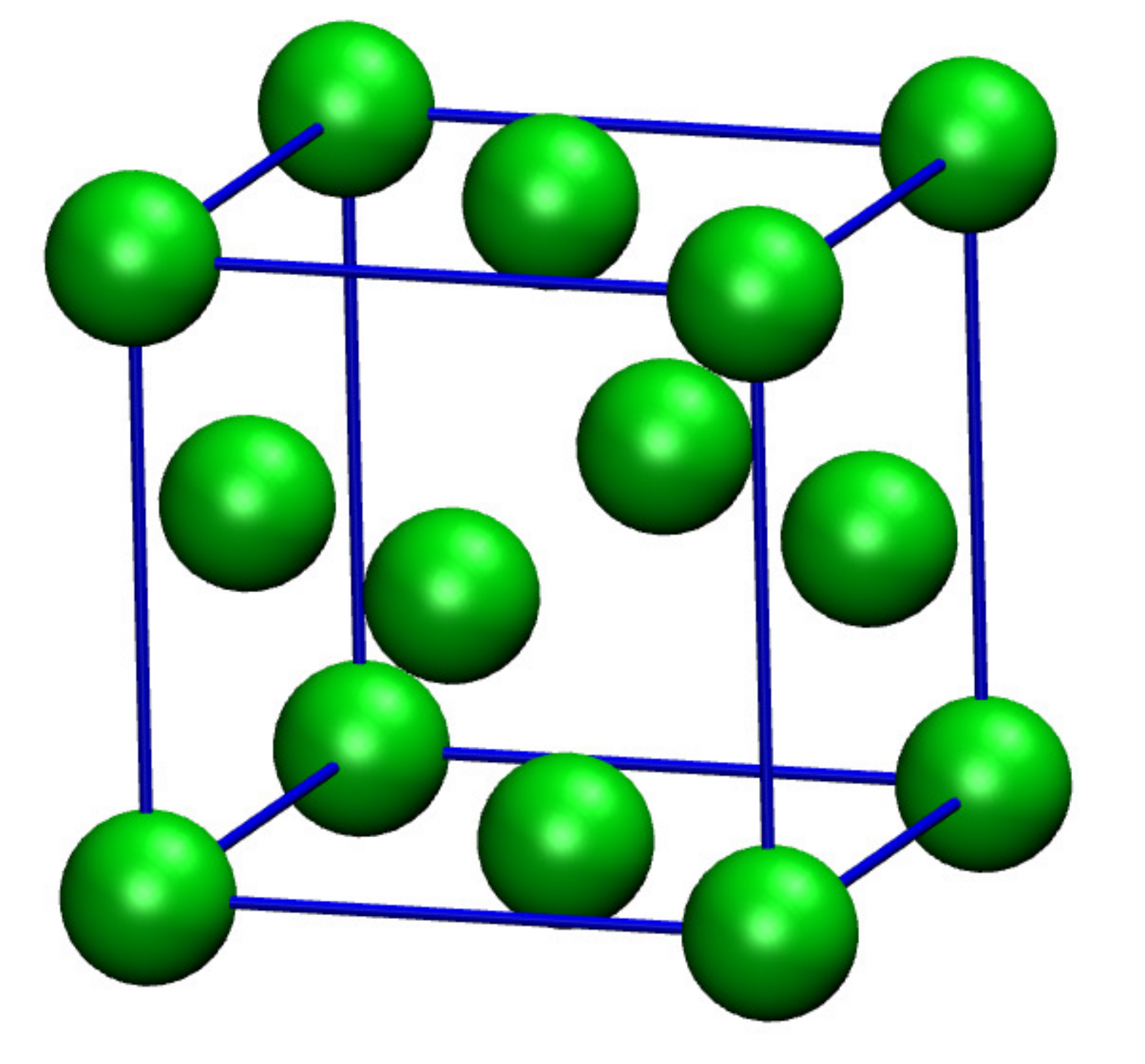}}
\subfigure[Octet-truss]
{\includegraphics[width=%
0.25\textwidth]{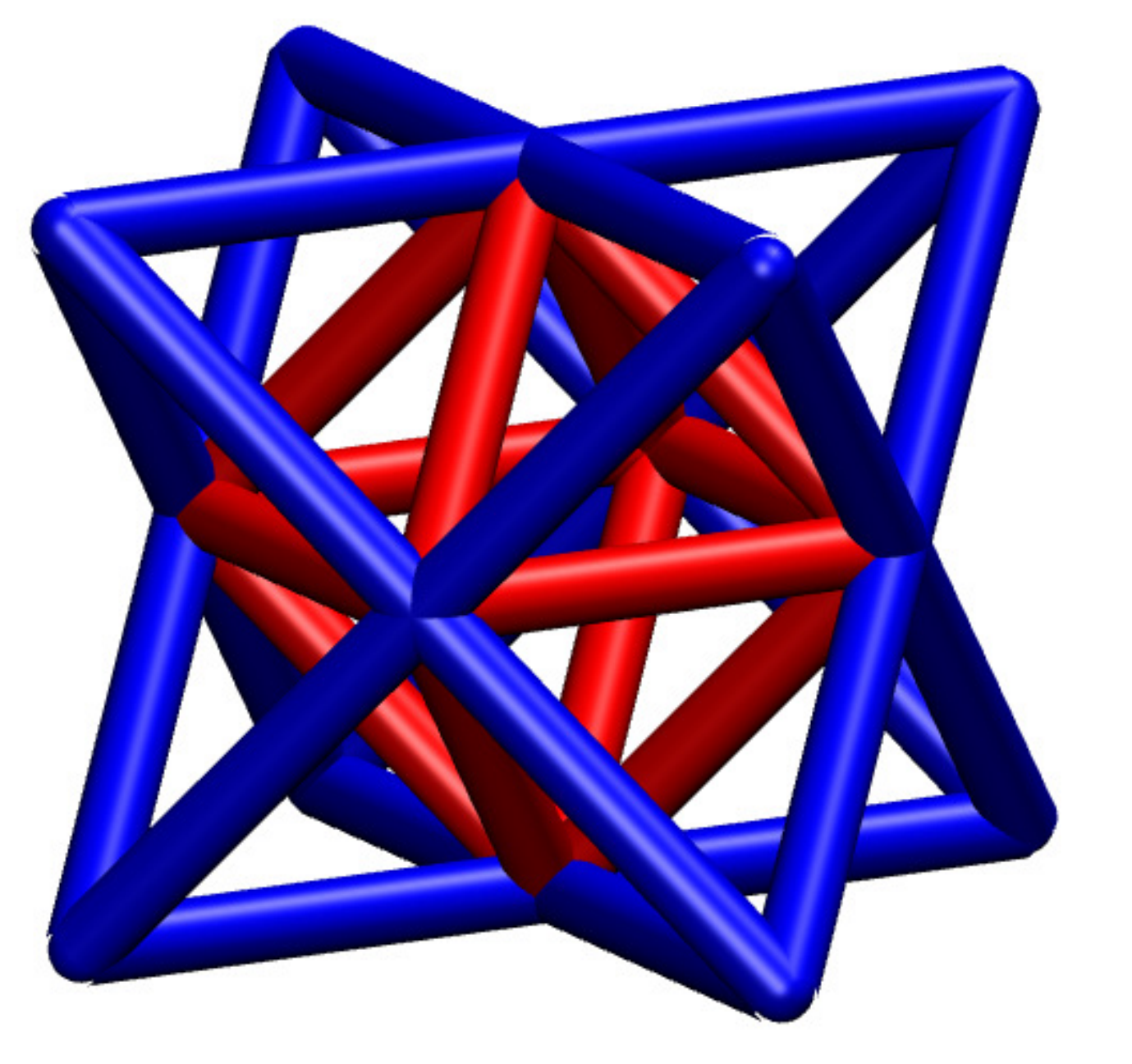}}
\subfigure[FCC unit cell + octet-truss]
{\includegraphics[width=%
0.25\textwidth]{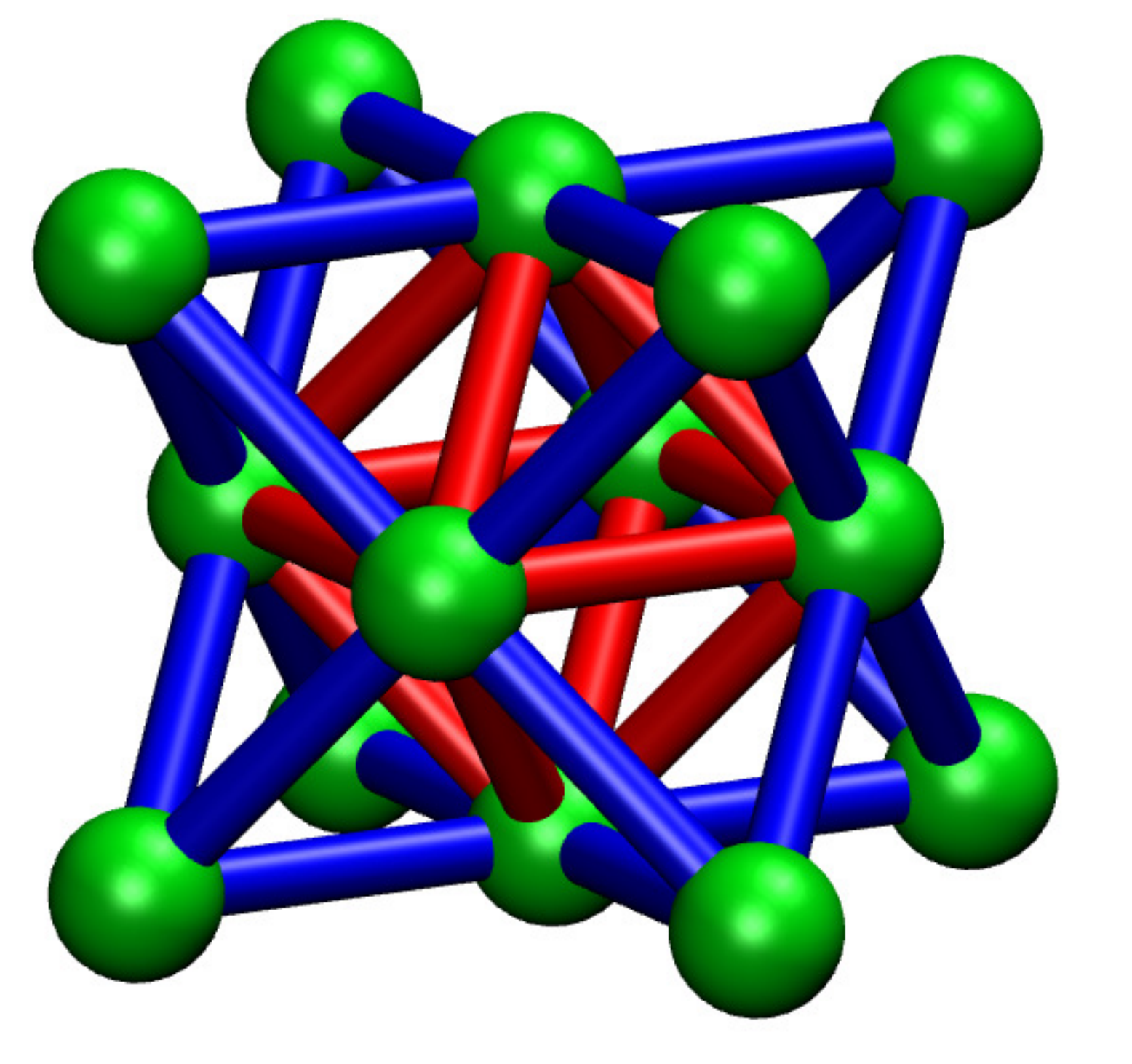}}
\caption{a) Unit cell with face-centered spheres. b) FCC points connected by sticks to give an octet-truss geometry. (c) Structure obtained by merging a) and b). Subsequently this last geometry was tiled by a graphene net.}
\label{fig:1}
\end{figure*}

Nevertheless, the realization of these periodic graphene 3D nanostructures
has been achieved for only few geometries due to the complexity of the
synthesis processes.
In this regard, one of the architectures of carbon-based materials that can
be most easily manufactured is based on the face-centered cubic (FCC) geometry
(Fig.~\ref{fig:1}a). Graphene nanostructures with this geometry have been
indeed synthesised, for example, by growing graphene on a FCC assembly of
silica nanoparticles \citep{Yoon2013}. Following this approach, a FCC
network of hollow graphene spheres in contact each other was
obtained. FCC carbon-based structures can also be realized by covering
micrometric 3D trusses with graphene using lithography \citep{Xiao2012}.

On the other hand, computer simulations can be used to
perform a detailed screening of different architectures and to help our
understanding of their specific properties. To achieve this goal, it is of course necessary to model
realistic structures and not only those ideally built from regular blocks
of fullerenes and nanotubes.

\begin{figure}[t]
\centering
\includegraphics[width=%
0.4\textwidth]{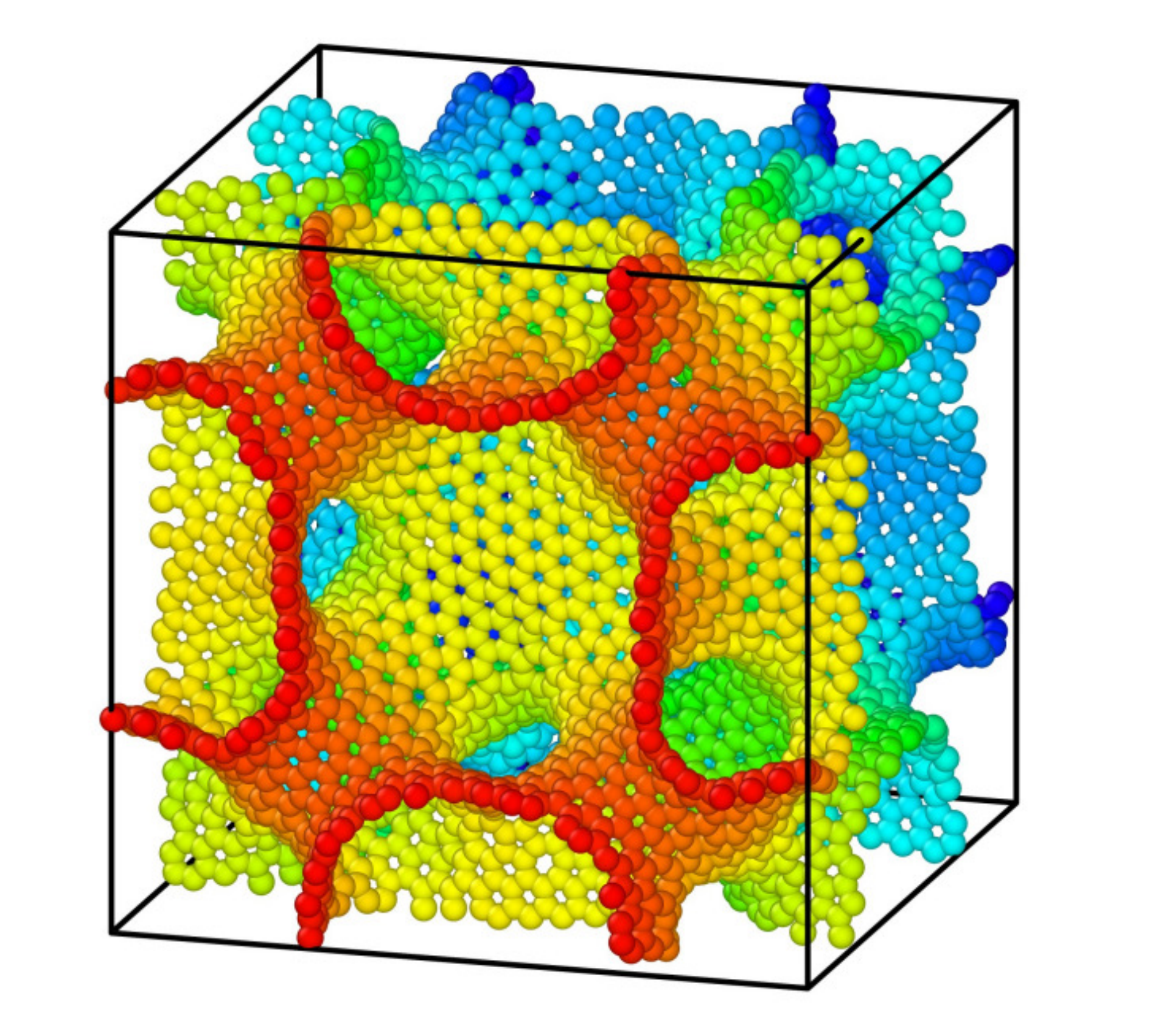}
\caption{FCC cell of the nanotruss. The faces of the cubic box are
  perpendicular to the [100], [010], [001] directions. In this figure, the
  nanotube diameter is 0.11 nm, the sphere diameter is 3.4 nm and the cube
  edge is 5.5 nm. Colors have been used for visualization purposes only and
  have no physical meaning.}
\label{fig:2}
\end{figure}

Here, we present the results of computer simulations of novel
graphene nanotruss architectures with FCC crystal structure
(Fig. \ref{fig:2}) investigating their mechanical properties with
atomistic resolution. Firstly, we compute the stiffness matrix. Subsequently, we perform simulations in
tensile and compressive regimes, studying the stress-strain curves as a
function of different geometries, with particular consideration to
assessing local instabilities and mechanical hysteresis. Furthermore,
observables characterising these FCC structures, such as Poisson ratio and
scaling relations between the Young modulus and the density, are determined
and compared to the performance of standard materials, such as
graphite. Finally, since graphene nanotrusses could be potentially suitable
for impact energy absorption and capable of propagating nanoscopic size
effects to macroscopic scales, we investigate their energy absorption
efficiency.

In designing a three-dimensional graphene-based nanomaterial, we combined ideas coming from both nanotruss network desing and open-cell foams~\cite{Ashby2006}. The former are micrometric structures built joining struts in a regular fashion (see Fig.~\ref{fig:1}b), whereas the
latter are porous materials in which pores are connected to each other.  
The periodicity of microtruss networks, particularly in their octet
configuration (see Fig.~\ref{fig:1}b), enables a good scaling of their
mechanical properties. In general, microtruss networks can be made by
hollow or solid struts, the former showing -- in general -- better elastic
properties. However, these materials are fragile and tend to break or
deform irreversibly under compression, the principal point of failure being
the connection between the struts.

In this paper, we explore the possibility of enhancing the properties of
truss networks by a modification of the way in which the struts are
connected. Taking inspiration from the structure of foams, we envisage
a nanostructured material in which hollow carbon nanotubes are connected to
spheres placed in FCC configurations. 
Similar materials have already been investigated in the case of
body-centered cubic geometry, with emphasis on the mechanical properties
under tension~\citep{Wu2013}.

\section{Modeling nanotruss geometries}

\begin{figure*}[]
\centering
\subfigure[Starting configuration]{
\includegraphics[width=0.30\textwidth]{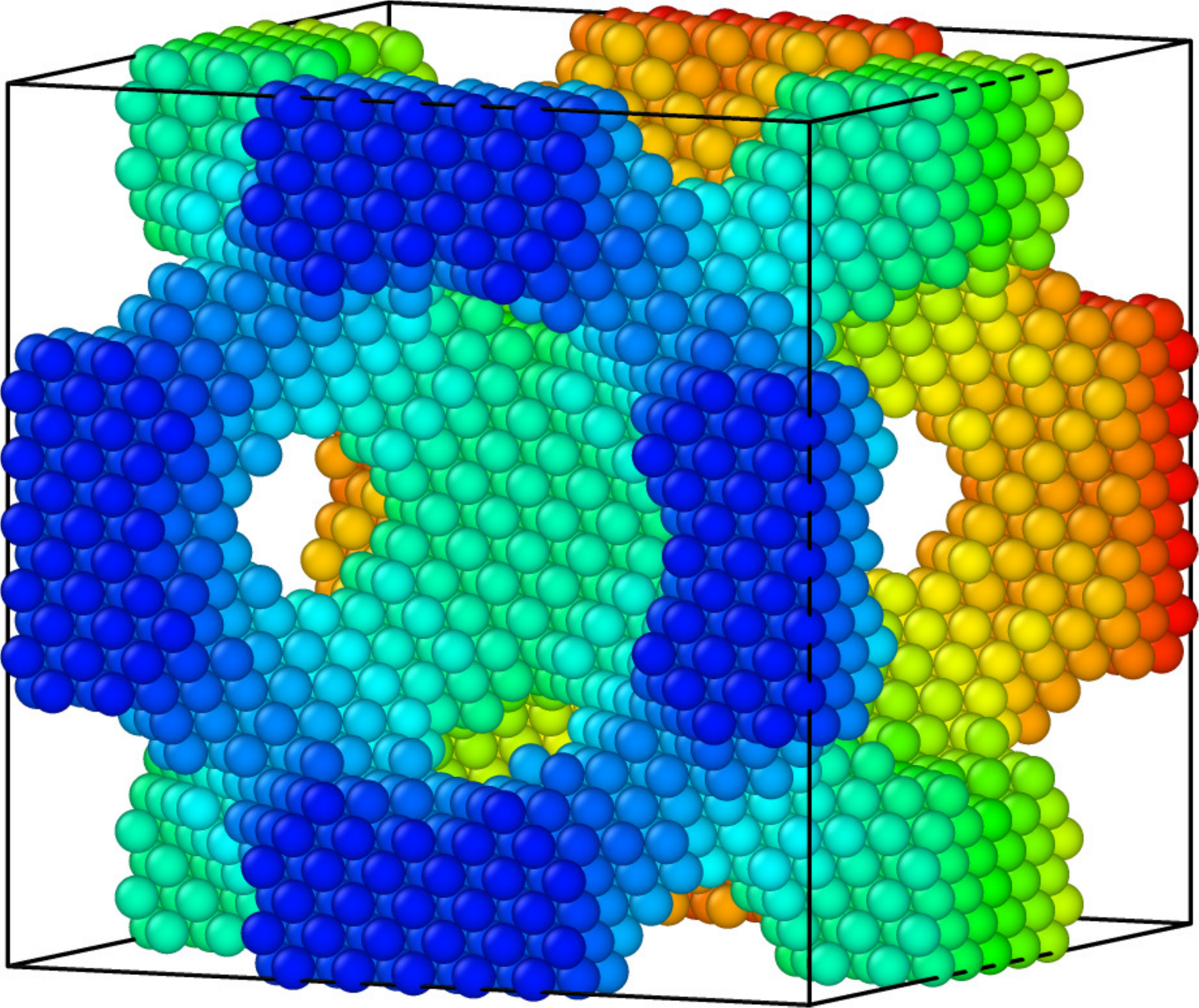}}
\subfigure[Relaxation of the points]{
\includegraphics[width=0.30\textwidth]{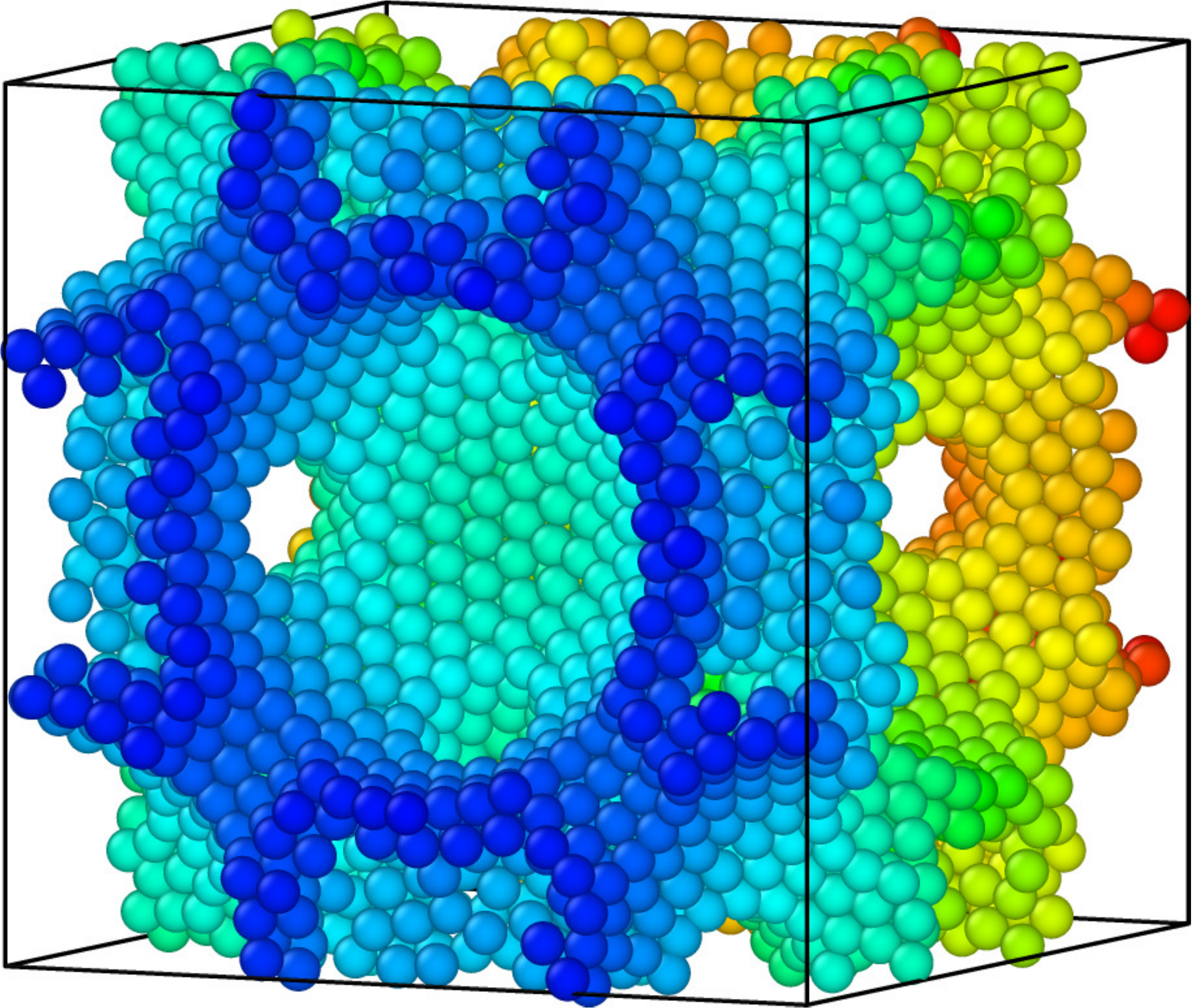}}
\subfigure[Deletion of the points in excess]{
\includegraphics[width=0.30\textwidth]{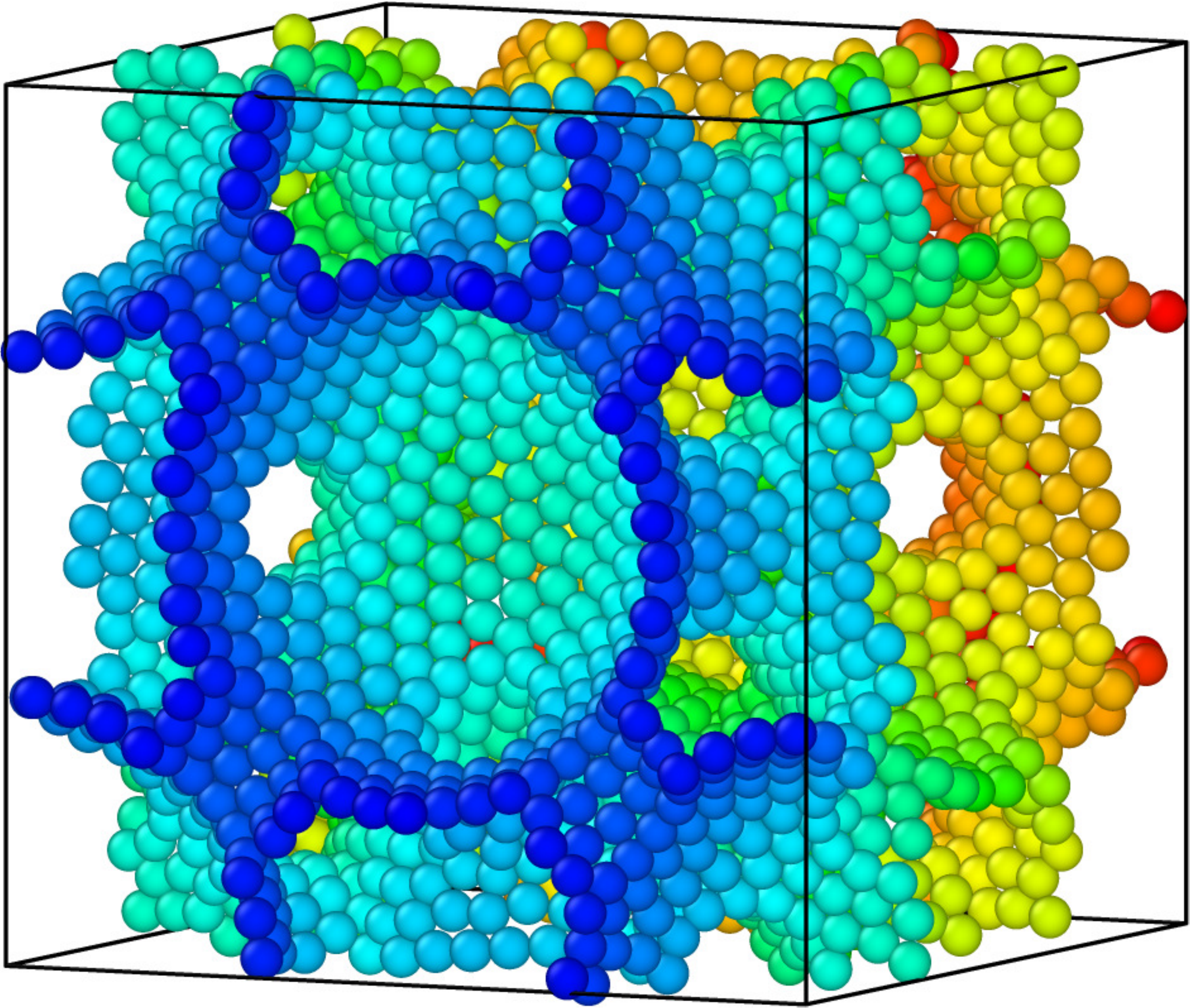}}
\caption{The three step sequence for triangulating the surface. Panel a)
  shows the initial condition in which points are arranged in a regular
  grid.  Subsequently, the position of the points is relaxed and they are
  attracted towards the surface by means of a molecular dynamics run (panel
  b). Finally, the points that do not belong to the first layer are deleted
  to avoid multilayer structures (c). The LJ net is ready for the Voronoi
  dualization, where it acquires a graphene-like net (see
  Fig. \ref{fig:2}). Colors have been used for visualization purposes
  only and have no physical meaning.}
\label{fig:3}
\end{figure*}

\begin{figure}[t]
\centering
\includegraphics[width=0.24\textwidth]{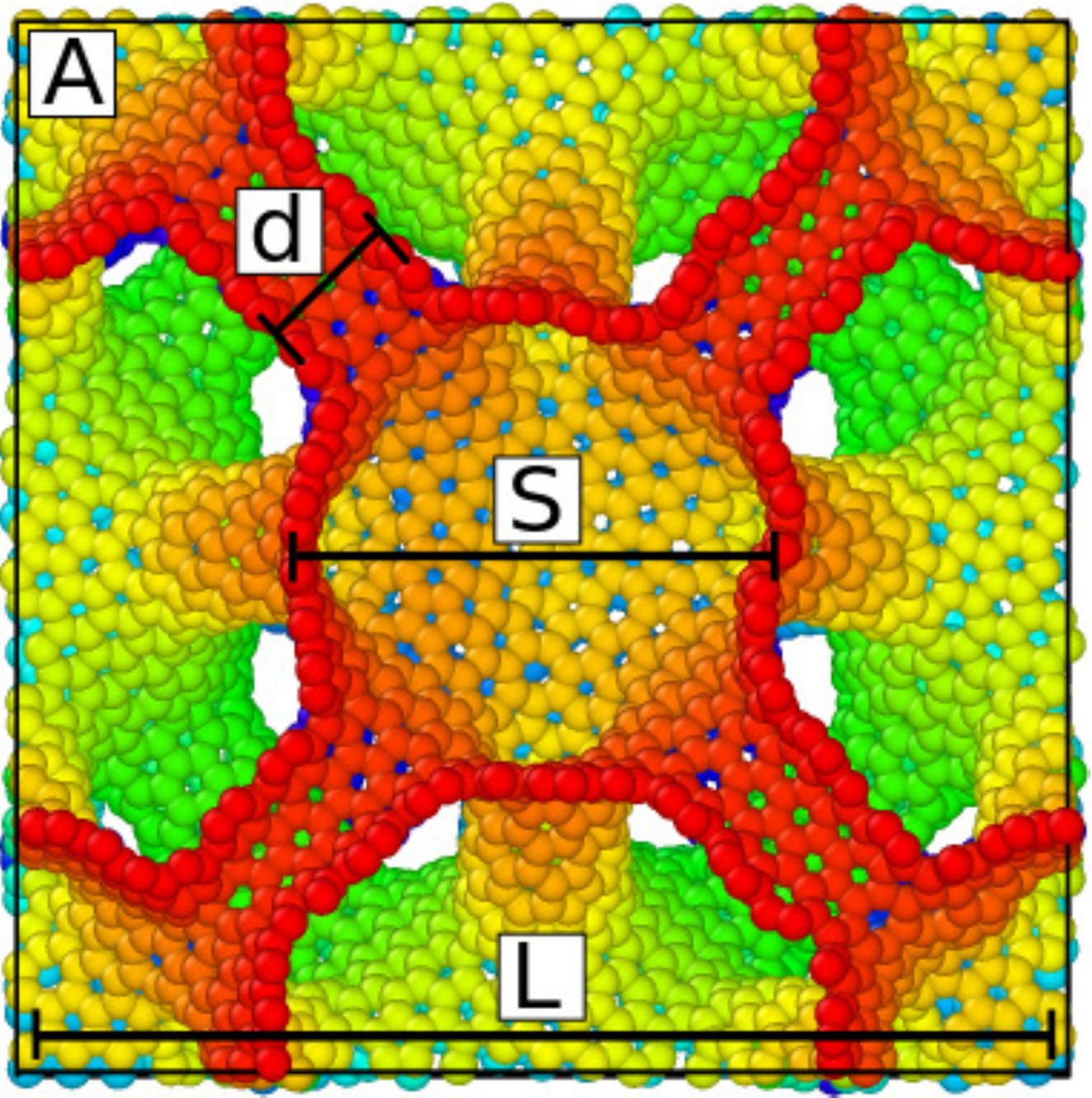}
\includegraphics[width=0.24\textwidth]{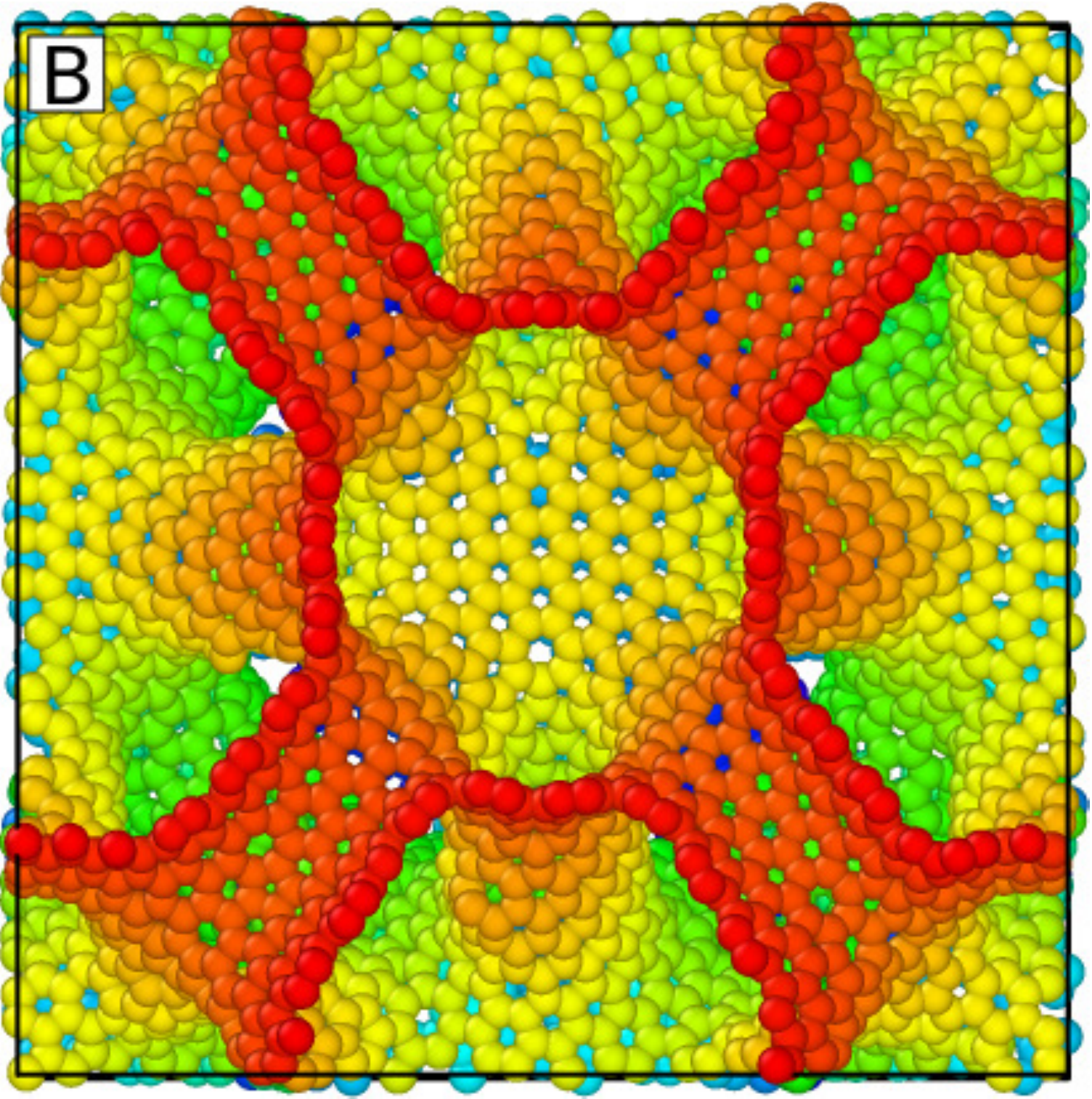}
\includegraphics[width=0.24\textwidth]{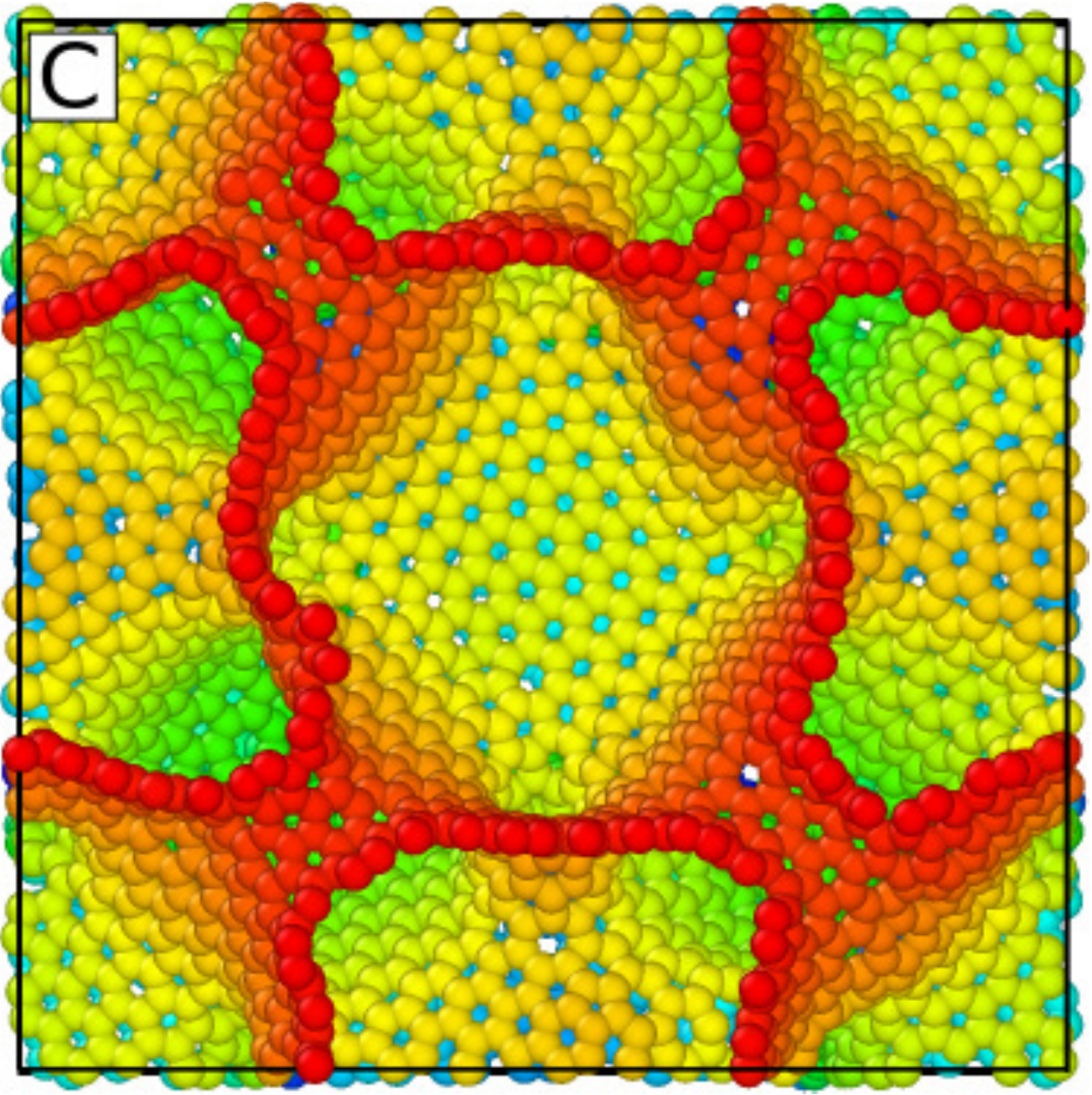}
\includegraphics[width=0.24\textwidth]{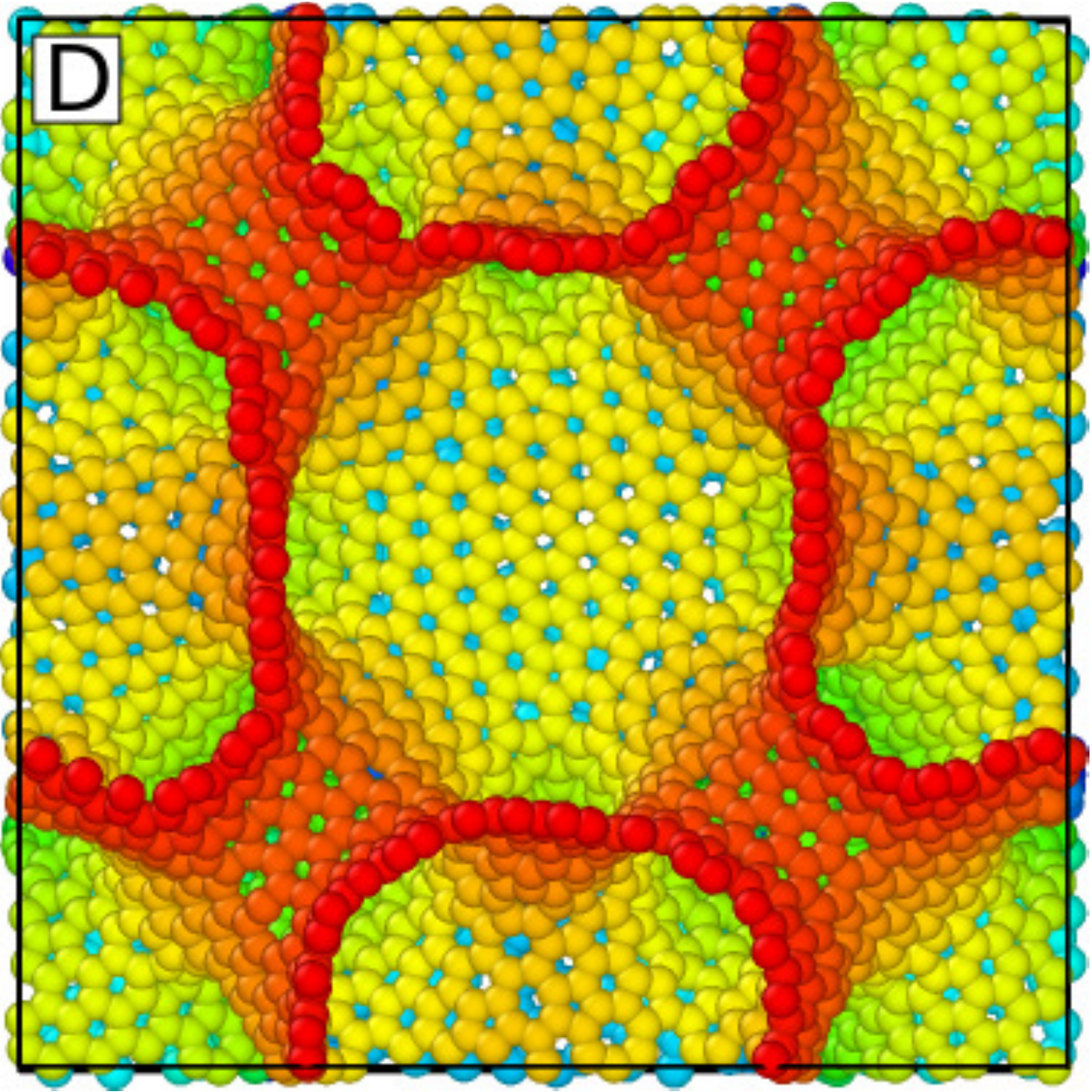}
\caption{Side views of the FCC cells. The cube faces are perpendicular to the [100], [010] and [001] directions. The diameter of the nanotubes is 0.8 nm for unit cells A and C, 0.11 nm for B and D. The sphere diameter is 2.7 nm for unit cells A and B and is 3.4 nm for unit cells C and D. The box edge is 5.5 nm for all four unit cells. Colors have been used for visualization purposes only and have no physical meaning.}
\label{fig:4}
\end{figure}

In order to build nanotruss networks, we imagine to cover with graphene a
surface made by FCC spheres (see Fig.~\ref{fig:1}a) joined by struts in
octet-truss geometry \citep{Deshpande2001} (see Fig.~\ref{fig:1}b). The
resulting surface is depicted in Fig.~\ref{fig:1}c.

Once the nanotruss structure is defined, we have to perform a surface
tessellation to create the actual carbon nanotruss. We achieve that by tiling the surface with regular triangles, and then using the Voronoi partitioning to dualize it \citep{Jung2015,Taioli2016}. The triangulation is
associated with a dense uniform packing, having a fixed lattice spacing,
chosen to be almost double the carbon-carbon distance of graphene. This procedure follows closely that one reported in Ref.\citep{Taioli2016}, differing in the way the initial triangulation of the surface is achieved. This triangulation was performed using the molecular-dynamics package LAMMPS~\citep{Plimpton1995}.

A number of points, enough to cover the entire surface, are distributed initially on a regular space-grid (Fig. \ref{fig:3}c) and interact each other via a
pair-wise Lennard-Jones (LJ) 12-6 potential, with potential parameters
($\sigma=3.2$~\AA, $\epsilon=2\times 10^{-4}$~eV) and a cutoff of
$3.2$~\AA, such that they behave almost like soft spheres. The parameter's choice is such that the points in the final configuration are distributed on the surface at a distance close to the
carbon-carbon bond length ${a}_{CC} = 1.42$~\AA.
The particles were attracted to the surface using a LJ 9-3 potential
($\sigma=2.0$~\AA, $\epsilon=1.0$~eV) with a cutoff of $10.0$~\AA\ between the
points and the surface itself.  After relaxation by means of a NVE
integration with a viscous damping force (Fig. \ref{fig:3}b), the particles
distant more than $0.3$~\AA\ from the surface were deleted to realize a
single layer structure (Fig. \ref{fig:3}c). During the whole procedure we
imposed periodic boundary conditions, as detailed below.

A number of defects, distributed all over the structure, appear after the
LJ triangulation. To generate nanotrusses that provide models for
sp$^2$-bonded carbon atoms in graphene, one needs to apply a topological
dualization (Voronoi partitioning) to the LJ optimized lattice. We initially
computed the adjacency matrix of each particle, where a neighbour was
defined as a particle closer than $\sqrt{3}\times {a}_{CC}$ . Distances
were evaluated in 3D space and not on the surface. 
As a final step, we took the Voronoi dual of the points triangulating the
surface, using a refining procedure~\citep{Taioli2016} to obtain a
configuration of carbon atoms containing only pentagonal, hexagonal and
heptagonal rings.
These configurations were further annealed by MD with AIREBO-type potentials to obtain the
optimized carbon nanotruss networks.

It is worthwhile to note that in evaluating the distances in 3D space the
Voronoi dualization automatically smooths possible steps at the
intersection between different parts of the surface (e.g. tube-sphere in
our case).  Heptagonal and pentagonal defects appear in these structures as
can be seen in Fig.~\ref{fig:2}, where we show a unit cell of a model
nanotruss obtained using this procedure.

To learn about possibly novel mechanical behaviors related to nanotruss geometry, we choose a geometry that presents some particular features. In principle an hollow nanotruss with high connectivity at the nodes (12 in our case) has no degrees of freedom presents mechanical failures mainly at the nodes, as mentioned before. To enhance the possibility of internal degrees of freedom we introduced the hollow spheres as nodes. This allows, depending on the spheres' size, to obtain a more rigid structure for small loads and various internal degrees of freedom as the load increases.

Nanotruss networks are uniquely defined by fixing the edge length of the cubic box $L$, the diameter of the nanotubes $d$ and the diameter of the spheres $S$. Four different structures, shown in Fig.~\ref{fig:4}, have been obtained by modifying these parameters, as reported in Tab.~\ref{tab:Table 1}. Using these carbon-based structures, we investigate the dependence of mechanical properties on sphere's and nanotube's diameters, and on cell orientation.

\begin{table}[t]
\centering
\small
\begin{tabular}{ccccc}
\toprule 
Unit cell      & Sphere  & Nanotube   & Box side  \\ 
            &   diameter      & diameter         &  length        \\
           & $S$ (nm)        & $d$ (nm)         &  $L$ (nm)         \\
\midrule
A  & 2.7 & 0.8 & 5.5 \\ 
B  & 2.7 & 1.1 & 5.5 \\ 
C  & 3.4 & 0.8 & 5.5 \\ 
D  & 3.4 & 1.1 & 5.5 \\ 
\bottomrule
\end{tabular}
\caption{Parameters used to build the four nanotruss unit cells reported in Fig.\ref{fig:4}.}
\label{tab:Table 1}
\end{table}

In order to study the mechanical properties of these structures as a function of cell orientation, the calculation supercells were obtained by replicating the unit cell of each structure according to the periodic boundary conditions along the [100] direction, rotated and finally truncated to obtain configurations with faces perpendicular to the [110] and [111] directions. All considered samples are composed of more than one unit cell to limit the influence of the boundary conditions on the possible reciprocal sliding of spheres' planes. In this regard, starting from four unit cells, we built four structures for each direction of sampling. The samples derived from the A, B, C and D unit cells reported in Tab.~\ref{tab:Table 1} will be numbered from 1 to 4 onwards.
The number of atoms in the samples is dependent on the direction, approximately $4\times10^4$ for the [100] direction, $2\times10^4$ for the [110] and $3\times10^4$ for the [111], respectively.
\\

\section{Computational methods}

Molecular dynamics simulations were carried out using LAMMPS \citep{Plimpton1995}. The carbon-carbon atomic interaction was modeled through the AIREBO potential \citep{Stuart2000}. Atomic configurations were visualized by using the OVITO package \citep{Stukowski2010} or VMD \citep{Humphrey1996}.

The samples were annealed to randomize the presence of defects within the calculation supercell. Samples were first heated at $3500$~K and then equilibrated at this temperature for $100$~ps. Finally, they were cooled down to $700$~K in $100$~ps using a viscous damping force. The annealing was performed using the standard value for the cutoff parameter for the REBO part of the potential and performed within the microcanonical ensemble (NVE).  

We started the mechanical characterization of the samples computing the stiffness matrix.
For anisotropic materials the stress and strain tensors are related by a fourth rank tensor, having 21 independent elements, as follows:

\begin{equation}
\sigma_{ij} =C_{ijkl} \varepsilon_{kl}
\end{equation} 
\\
where $C$ is named \emph{stiffness tensor}.
In the particular case of cubic symmetry the stiffness tensor has only three terms and the linear system of equations can be written explicitly, as follows:

\begin{equation}
  \small
  \begin{pmatrix}
    \sigma_{xx} \\
    \sigma_{yy} \\
    \sigma_{zz} \\
    \sigma_{yz} \\
    \sigma_{zx} \\
    \sigma_{xy} \\
  \end{pmatrix} 
  =
  \begin{pmatrix}
    C_{11} & C_{12}   & C_{12}   &    0       &   0      &    0      \\
    C_{12} & C_{11}   & C_{12}   &    0       &   0      &    0      \\
    C_{12} & C_{12}   & C_{11}   &    0       &   0      &    0      \\
     0     &  0       &    0     &   C_{44}   &   0      &    0      \\
     0     &  0       &    0     &    0       & C_{44}   &    0      \\
     0     &  0       &    0     &    0       &   0      &  C_{44}
  \end{pmatrix} 
    \begin{pmatrix}
    \varepsilon_{xx} \\
    \varepsilon_{yy} \\
    \varepsilon_{zz} \\
    2\varepsilon_{yz} \\
    2\varepsilon_{zx} \\
    2\varepsilon_{xy} \\
  \end{pmatrix} 
\end{equation}
\\
where the matrix is named \emph{stiffness matrix}.

With regard to the simulations in compressive and tensile regime, all samples were equilibrated at zero pressure and $1$~K temperature with Nosé–Hoover barostat and thermostat. The adaptive cutoff parameter of the potential has been set to 2.0 Å to better describe the near-fracture regime \citep{Shenderova2000}. The equations of motion were solved with the velocity-Verlet integration method using a time step of $1$~fs. Mechanical properties were assessed in the isobaric-isothermal ensemble (NPT), adding a drag term to smooth out the pressure oscillations. The uni-axial tensile strain was applied up to the sample fracture in each case. 

\begin{figure}[]
\centering
\includegraphics[width=0.5\textwidth]{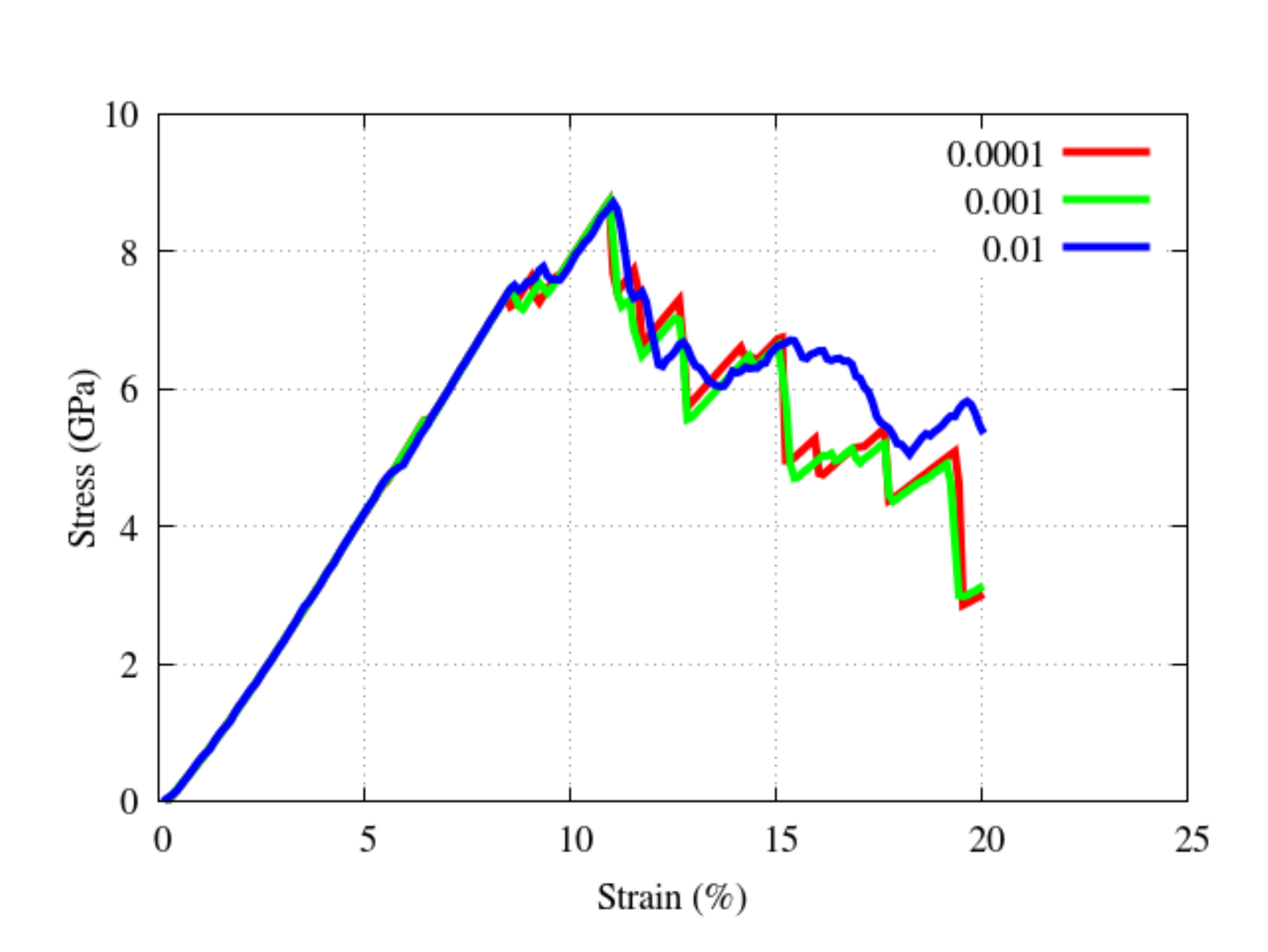}
\caption{Stress-strain curve dependence on strain rate in ps${^{-1}}$ for the [100] unit cell 4.}
\label{fig:5}
\end{figure}

The engineering strain parallel to the direction of deformation is defined as
\begin{equation}
\varepsilon = \frac{L-L_0}{L} = \frac{\Delta L}{L}
\end{equation} 
where $L_0$ and $L$ are the starting and current length of the 
sample in the direction of loading. To determine the stress, the pressure stress tensor components in response to the
external deformation are computed by
\begin{equation}\label{pressure}
P_{ij} = \frac{\sum_k^N{m_k v_{k_i} v_{k_j}}}{V}+ \frac{\sum_k^N{r_{k_i} f_{k_j}}}{V}
\end{equation} 
where $i$ and $j$ label the coordinates $x$, $y$, $z$; $k$ runs over the
atoms; $m_k$ and $v_k$ are the mass and velocity of $k$-th atom; $r_{k_i}$ is the position of $k$-th atom; $f_{k_j}$ is the force $k$-th atom; and, finally, $V$ is the volume of the simulation box.
The pressure in equation \ref{pressure} includes both kinetic energy (temperature) and virial terms, as sum of atomic pairs. The computed stress is the \emph{true stress} because the pressure is measured with respect to the instantaneous section area of the samples. The uni-axial compressive strain was applied up to reaching $25$~\% and $50$~\% total strain. The applied strain rate is chosen equal to $0.001$~ps$^{-1}$, such to converge the Young modulus and tensile strength, as shown in Fig.~\ref{fig:5}. Stress and strain were saved every 1000 time steps.\\

The observables that we calculate to characterize the mechanical properties of the nanotrusses are, in addition to the stiffness tensor, the Young modulus, fracture stress and fracture strain. The toughness is also evaluated as the area under the stress-strain curve up to the fracture stress. Indeed, the samples have no plastic deformation but several sequential fractures.
Stress-strain curves of carbon nanotrusses present a not fully linear behaviour (Fig. \ref{fig:5}). Thus, the definition of one only slope does not guarantee an accurate fit of this curve and we are forced to introduce two different values of the Young modulus to characterize the mechanical
behaviour. In particular, the first value of the Young modulus is obtained as the tangent at zero strain, while the second one from a linear fit between 5\% to 8\% strain.

We also performed the calculation of the Poisson ratio $\nu$, defined as the negative ratio between the transverse deformation $\varepsilon_{\mathrm{T}}$ and the longitudinal one $\varepsilon_{\mathrm{L}}$:
\begin{equation}
\nu =- \frac{\varepsilon_{\mathrm{T}}}{\varepsilon_{\mathrm{L}}}
\end{equation} 

Here we extend the concept of Poisson ratio to deformations beyond the linear regime, and use it to quantify the lateral deformation of the material. A similar extension is done for the Young modulus.

\section{Results and discussion}
\subsection{Stiffness matrix}
The elastic properties of the nanotrusses were first assessed by computing the stiffness matrix. The outcome of this computation will be useful also to check if the choice of the structures represents a realistic model of these materials. Indeed, the stiffness matrix of realistic systems can be used to measure the degree of anisotropy by comparison with a perfect FCC cubic cell.
The nanotruss networks under investigations present a face centered cubic geometry that would influence the stiffness matrix.
The computed stiffness matrix for the sample 3, with faces perpendicular to [100], [010] and [001] directions, is:

\begin{equation*}
  \begin{pmatrix}
    63.0  & 12.3   & 13.2  &  -0.43   &    0.19   &    -0.92     \\
      &    58.9    & 12.4  &  -0.28   &    0.28   &    -1.14     \\
      &            & 58.5  &   0.59   &    0.97   &     0.32    \\
      &            &       &   14.0   &    0.11  &      0.71     \\
      &     Sym    &       &          &    14.6   &     -0.40     \\
      &            &       &          &           &      13.9
  \end{pmatrix} 
\end{equation*}
\\
where the values are reported in GPa.
  
The matching with the cubic material stiffness matrix in Eq. 2 is not perfect, as small non-zero terms appear in the upper right part of the matrix. Additionally, terms that must be in principle equal show not negligible discrepancies. The reason of this anisotropy is principally due to the choice of the cell dimension used for the Voronoi tessellation of the surface. Applying the dualization to a single unit cell, the graphene net has to be periodic over a distance of the box edge $L$ and the room available to accommodate the defects is quite small.

\subsection{Tension}

\begin{figure}[]
\centering
\includegraphics[width=0.6\textwidth]{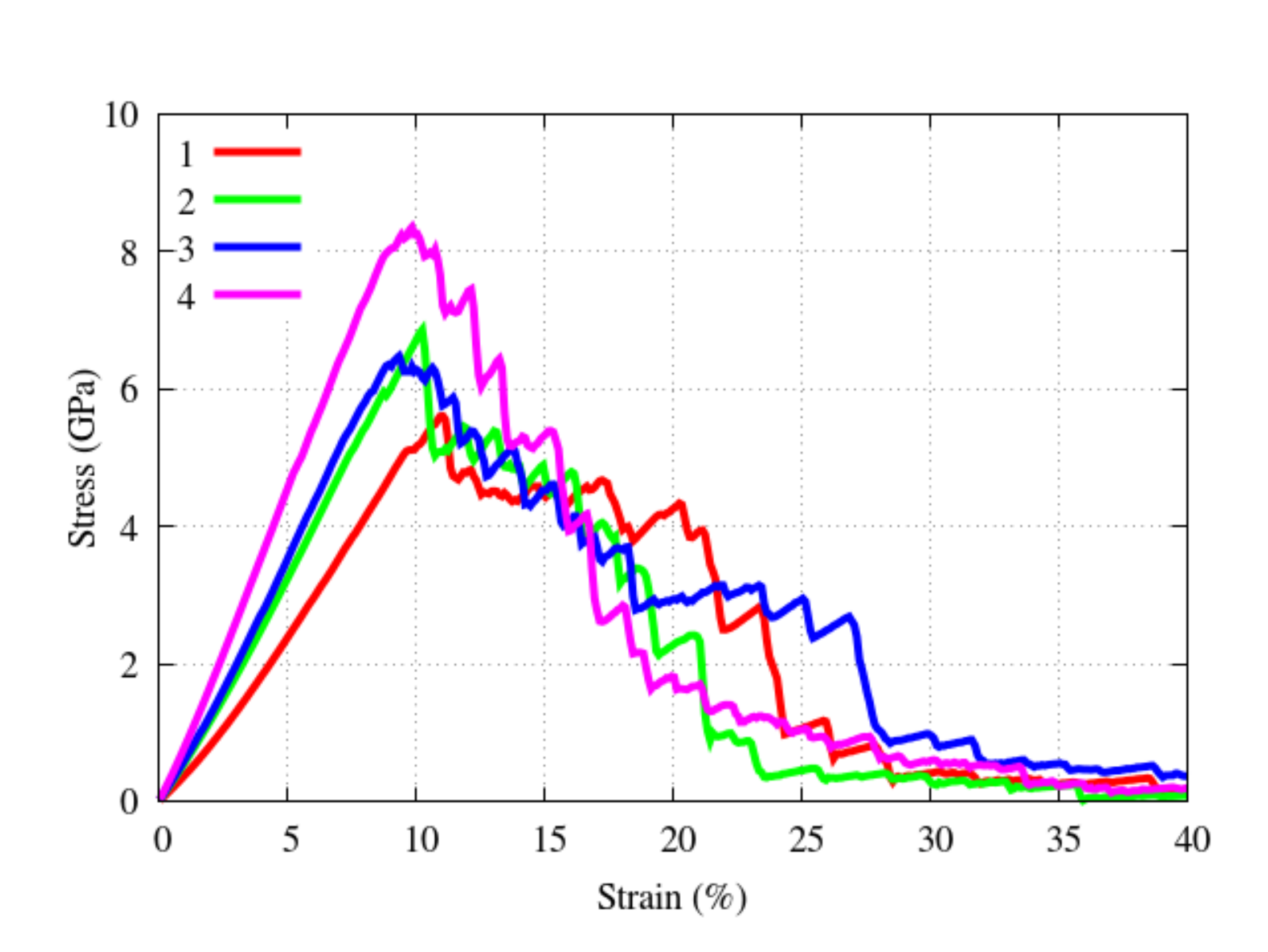}
\caption{Stress-strain curves of the four nanotruss networks under uni-axial tension along the [100] direction. The four curves present a Young modulus roughly proportional to tensile strength and a fracture strain of about $10\%$. Because the crack propagation can be strongly influenced by the size of the simulation box and periodic boundary conditions, the part of the curves beyond the maximum of stress should not be taken into account for a physical interpretation without the study of possibly box-size-dependent behavior.}
\label{fig:6}
\end{figure}

\begin{figure*}[]
\centering
\subfigure[0\% strain]{
\includegraphics[width=0.30\textwidth]{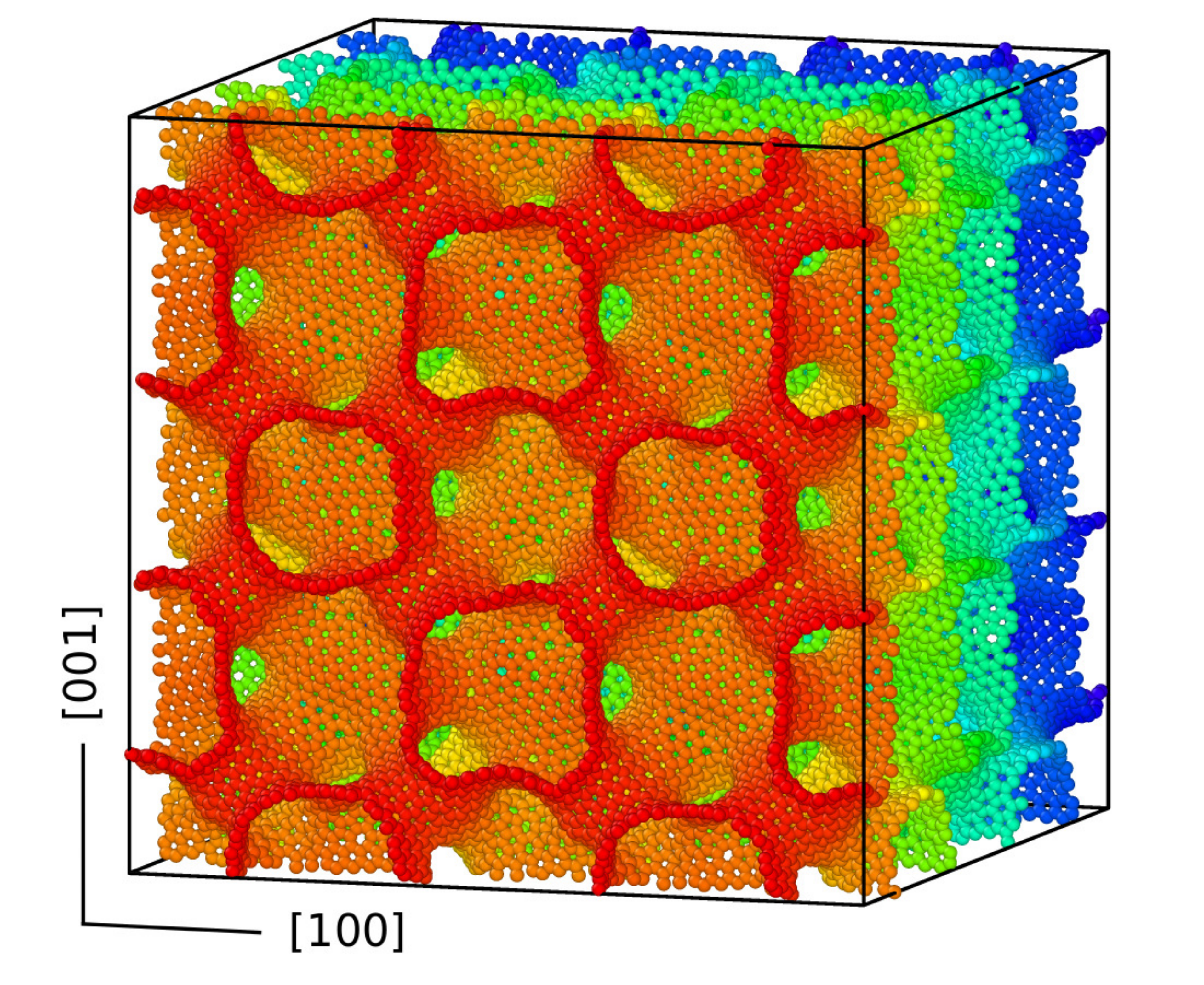}}
\subfigure[10\% strain]{
\includegraphics[width=0.30\textwidth]{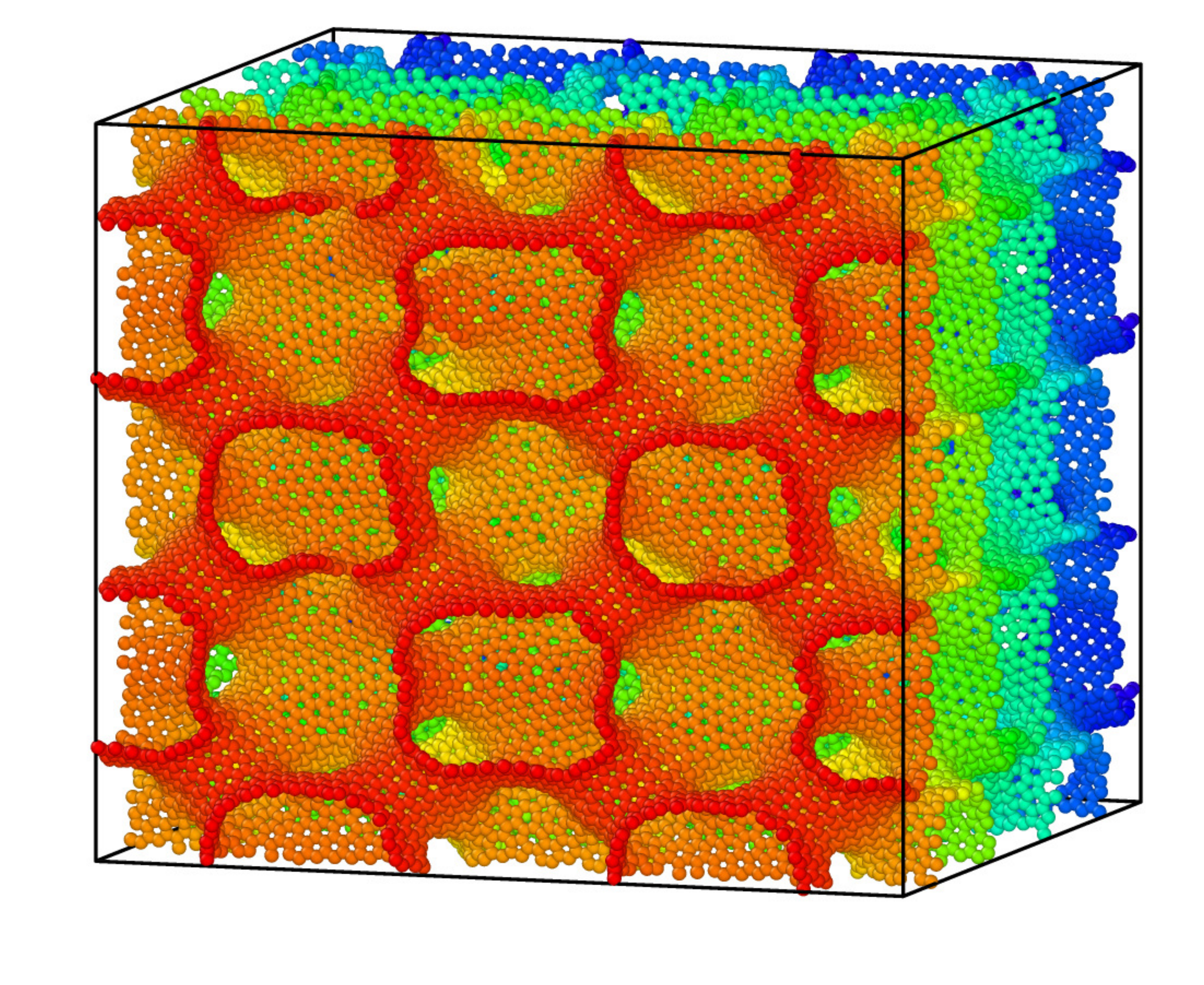}}
\subfigure[20\% strain]{
\includegraphics[width=0.30\textwidth]{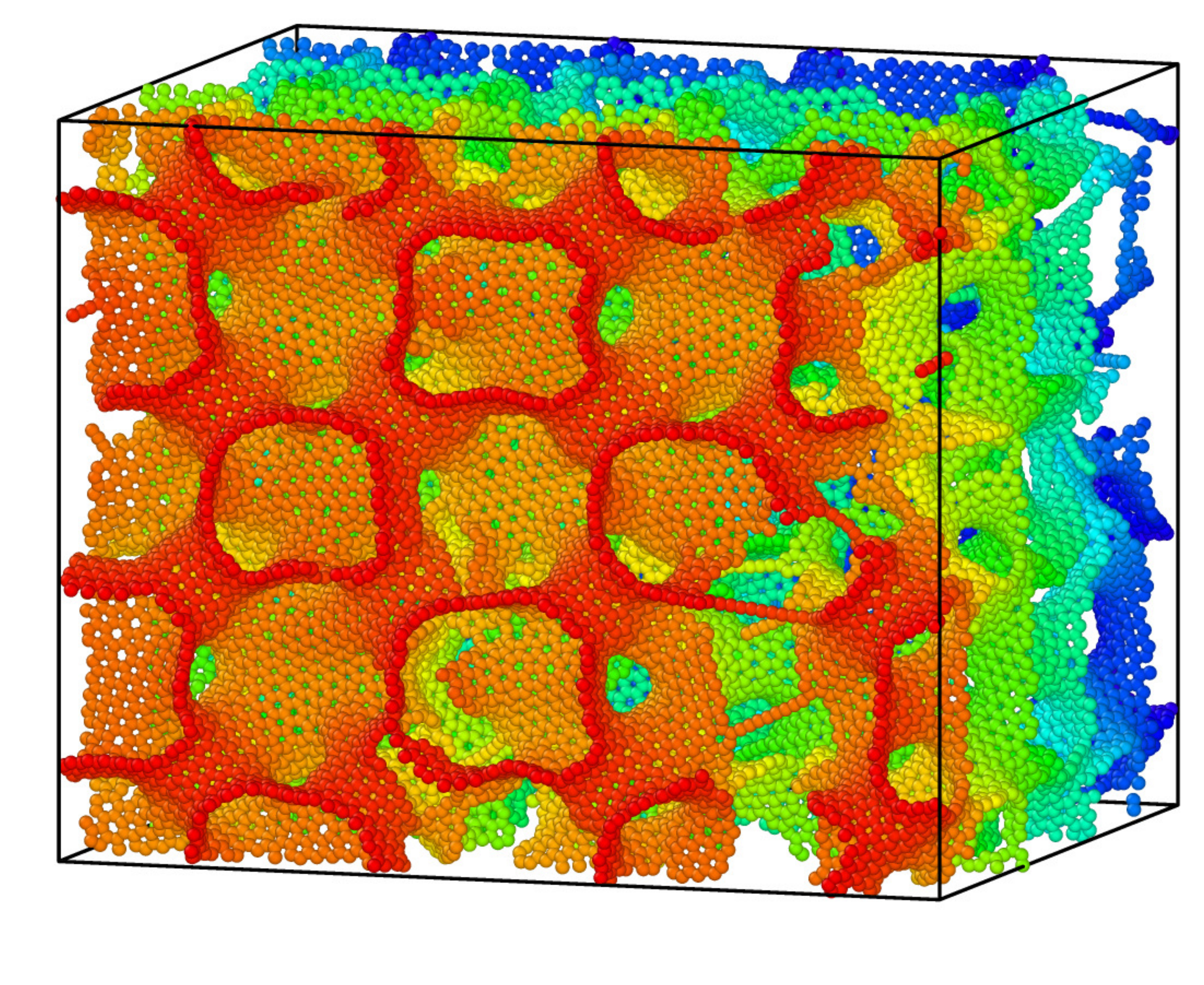}}
\caption{Images of sample 4 under traction in the [100] direction. The sample is built by replicating the unit cell D in the space building up a $2 \times 2 \times 2$ unit cell (a). The deformation is applied in the [100] direction. At 10\% strain the sample is near to the maximum stress and failure strain where cracks start to grow (b). In image (c), at 20\% strain, cracks are larger and there is not only one fracture but cracks formed in many positions. Colors have been used for visualization purposes only and have no physical meaning.}
\label{fig:7}
\end{figure*}

The mechanical properties of nanotruss networks, such as Young modulus, fracture strain and tensile stress, were further investigated via the assessment of their stress-strain curves. In Fig.~\ref{fig:6}, we report the stress-strain curve for the four samples along the direction [100] and in Fig.~\ref{fig:7} three snapshots of the sample 4 under tension.

The stress-strain curves show a typical elastic behavior for small deformations up to the tensile strength (snapshots a and b in Fig.~\ref{fig:7}), followed by a descending part that corresponds to the fracture of the samples (snapshot c in Fig.~\ref{fig:7}). The absence of a plastic plateau and instead the presence of a sharp maximum in the stress means that nanotruss networks display the brittle nature of the parent material, i.e. graphene.

In this study we do not attempt to extract informations from the part of the stress-strain curves beyond the fracture strain. Indeed, at variance with the Young modulus or specific toughness, the crack propagation can be strongly influenced by the size of the simulation box and periodic boundary conditions.

All the curves present a fracture strain of about $10\%$ so that the tensile strength is roughly proportional to the Young modulus. The network with the largest nanotube and sphere diameters (4) has larger Young modulus and tensile strength than the number 1, which present the smallest values (Fig.~\ref{fig:6}). The other two cases give comparable results. A similar response is found for the [110], and [111] directions. 

The values of Young modulus, tensile strength and fracture strain are reported in Tab.~\ref{tab:Table 2}. We also report the values for graphene as calculated using the AIREBO potential (cutoff set to $2.0$~Å) in Ref. \citep{Zhao2009}. We see an increasing of Young modulus and tensile strength for evidently more optimized structures, passing from sample 1 to sample 4. The shape and dimension of the nodes play an important role in the deformation of hollow truss networks, as recently shown for microtruss networks \citep{Schaedler2011, Valdevit2013}.

The stiffening behavior (i.e. parabolic shape of stress-strain curve in elastic regime) is due to the realignment and bending of the struts at smaller strain, and the stretching-dominated deformation of the realigned struts at larger strain, as noted in Ref. \citep{Wu2013} for similar structures. In this regard we can study the stiffening using the ratio between Young modulus 2 and 1, which takes the highest value for the sample 1 (1.61). Samples 2 a 3 give comparable results (1.37, 1.33). For the last sample (4) we find the lowest value (1.19). We underline the similarity between the behavior of the samples 2 and 3 in tension in contrast to what it will turn out in the compressive case. In fact, they have different geometries but almost the same mass density.

\begin{table*}[t]
\centering
\small
\begin{tabular}{ccccccc}
\toprule
             & Young   & Young   & Tensile   & Fracture  \\ 
    Sample   & modulus 1 & modulus 2 &  strength  &  strain     \\ 
             &(GPa)  &   (GPa)  &   (GPa)     &     (\%)    \\
\midrule
1 & 36 & 58 & 5.6 &   11.1 \\ 
2 & 54 & 74 & 6.8 &   10.2 \\ 
3 & 58 & 77 & 6.5 &    9.3 \\ 
4 & 79 & 92 & 8.3 &    9.8 \\ 
\midrule
Graphene & 10$^{3}$  & 10$^{3}$  & 98.5 &  16.5 \\
\bottomrule
\end{tabular}
\caption{Young modulus, ultimate strength, and fracture strain of
the various systems studied under tension in the [100] direction. These values are compared with the average of those reported for armchair and zigzag graphene using AIREBO potential \citep{Zhao2009}.}
\label{tab:Table 2}
\end{table*}

\begin{table*}[t]
\centering
\small
\begin{tabular}{ccccccc}
\toprule
           &  Density   & Specific & Specific  & Specific  &  Specific\\ 
 Sample    &            & modulus 1 & modulus 2 & strength  & toughness\\ 
           &   kg m$^{-3}$ & (MNm kg$^{-1}$) & (MNm kg$^{-1}$) & (MNm kg$^{-1}$) &     (MJ~kg$^{-1}$) \\
\midrule
1 &  634 & 57 & 91 & 8.8 & 0.5\\ 
2 &  707 & 76 & 105 & 9.6 & 0.5\\ 
3 &  705 & 82 & 109 & 9.2 & 0.4\\ 
4 &  692 & 114 & 133 & 12.0 & 0.6\\ 
\midrule
Graphite & 2250 & 444.4  & 444.4 & 43.8 & - \\
\bottomrule
\end{tabular}
\caption{Density, specific modulus and specific strength of
the four samples studied under [100] direction tensile strain. Specific toughness is calculated as the area under the stress-strain curve up to fracture strain per mass density.}
\label{tab:Table 2b}
\end{table*}

In Tab.~\ref{tab:Table 2b}  we report the specific modulus and the specific strength, which define the values per mass density. Furthermore, in the same table is reported the specific toughness, calculated as the total area under the stress-strain curves up to fracture strain, per mass density. By comparing Tabs.~\ref{tab:Table 2} and ~\ref{tab:Table 2b}, we note that even though the Young modulus and tensile strength for the nanotrusses are respectively two and one order of magnitude smaller than those reported for graphene, we obtain a different scenario when considering the specific strength and the specific modulus. The density of these nanotruss networks is about one third of that of graphite, thus enhancing the specific mechanical properties of these materials with respect to graphite. 

Considering Fig.~\ref{fig:7}, we note that cracks propagate from defective sites as, for example, a pair of heptagons.
The presence of defective sites is responsible for the small value of the fracture strain, compared to graphene. Furthermore, for defected carbon nanotrusses, as those studied in this work, crack propagation is diffuse, differently from ideal materials. Indeed, for ideal nanotube-fullerene networks cracks preferably are localized at the nanotubes-fullerenes junctions \citep{Wu2013}.

In the next section we will focus on the compressive regime in which a geometry-dependent behavior emerges more strongly than in the tensile one. 

\subsection{Compression}
\begin{figure}[t]  
\centering
\includegraphics[width=0.6\textwidth]{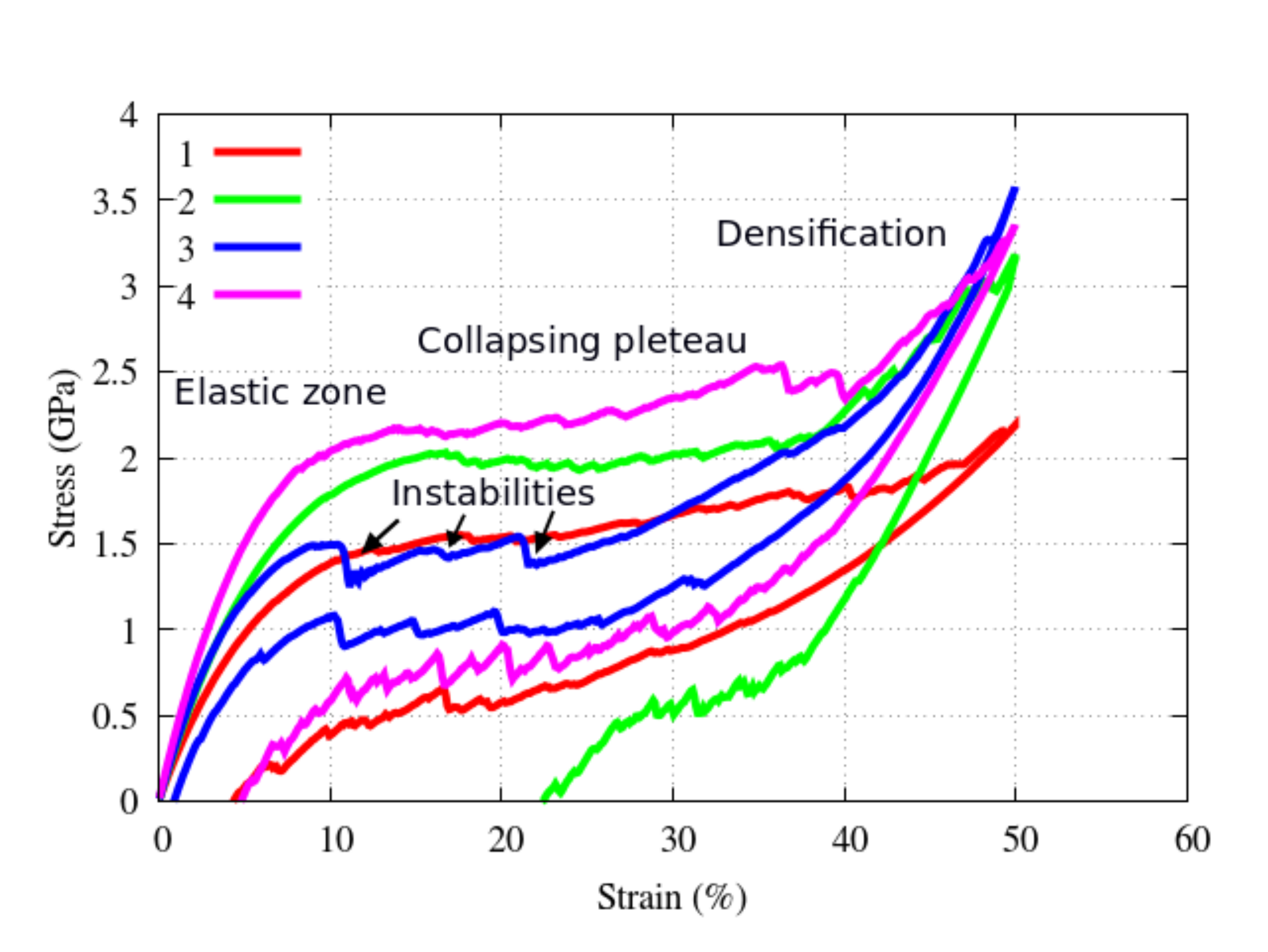}
\caption{Stress-strain curves of the four nanotruss networks under uni-axial compression along the [111] direction up to 50\% strain.}
\label{fig:8}
\end{figure}

In this section we present the results obtained for the samples under compressive load. At variance with the tension case, the compressive response is strongly related to the nanotruss geometry as well as to the direction of compression. In Fig.~\ref{fig:8} we report the stress-strain curve in compression along the [111] direction, which presents the most interesting features. The deformation reaches $50$\% strain for the largest compression, for which the main features of the compressive response are shown. 

Indeed, observing Fig.~\ref{fig:8}, the stress-strain curves are basically characterized by three regimes, typically found in foams and energy absorbing materials. At small strain we are in elastic regime, and the material is characterized by a full recovery of the original shape when the load is removed. Subsequently we observe a plateau, representative of the sample collapse at a nearly constant stress, by buckling or fracture of the building blocks (nanotubes and graphene spheres). Finally, one finds a steep rump in the stress-strain curve, representing full collapse or a densification regime of the structures.

\begin{table*}[h]
\centering
\small
\begin{tabular}{cccccc}
\toprule
             & Young   &  Plateau stress  &  Residual     & Residual   \\ 
    Sample   & modulus &   between $10$-$40$\%   &  deformation  &  deformation  \\ 
             &(GPa)    &   strain (GPa)    &  $25$\% (\%) &  $50$\% (\%)   \\
\midrule
1 & 27  & 1.6 & 0.6   &  4  \\ 
2 & 33  & 2.0 & 1.5   &  22 \\ 
3 & 36  & 1.6 & 0.2   &  1  \\ 
4 & 42  & 2.3 & 0.4   &  5  \\ 
\bottomrule
\end{tabular}
\caption{Young modulus, plateau stress, and residual deformation of
the four samples under uni-axial compression in the [111] direction.}
\label{tab:Table Comp}
\end{table*}

\begin{figure*}[t] 
\centering
\subfigure[Strain 0\%]{
\includegraphics[width=0.30\textwidth]{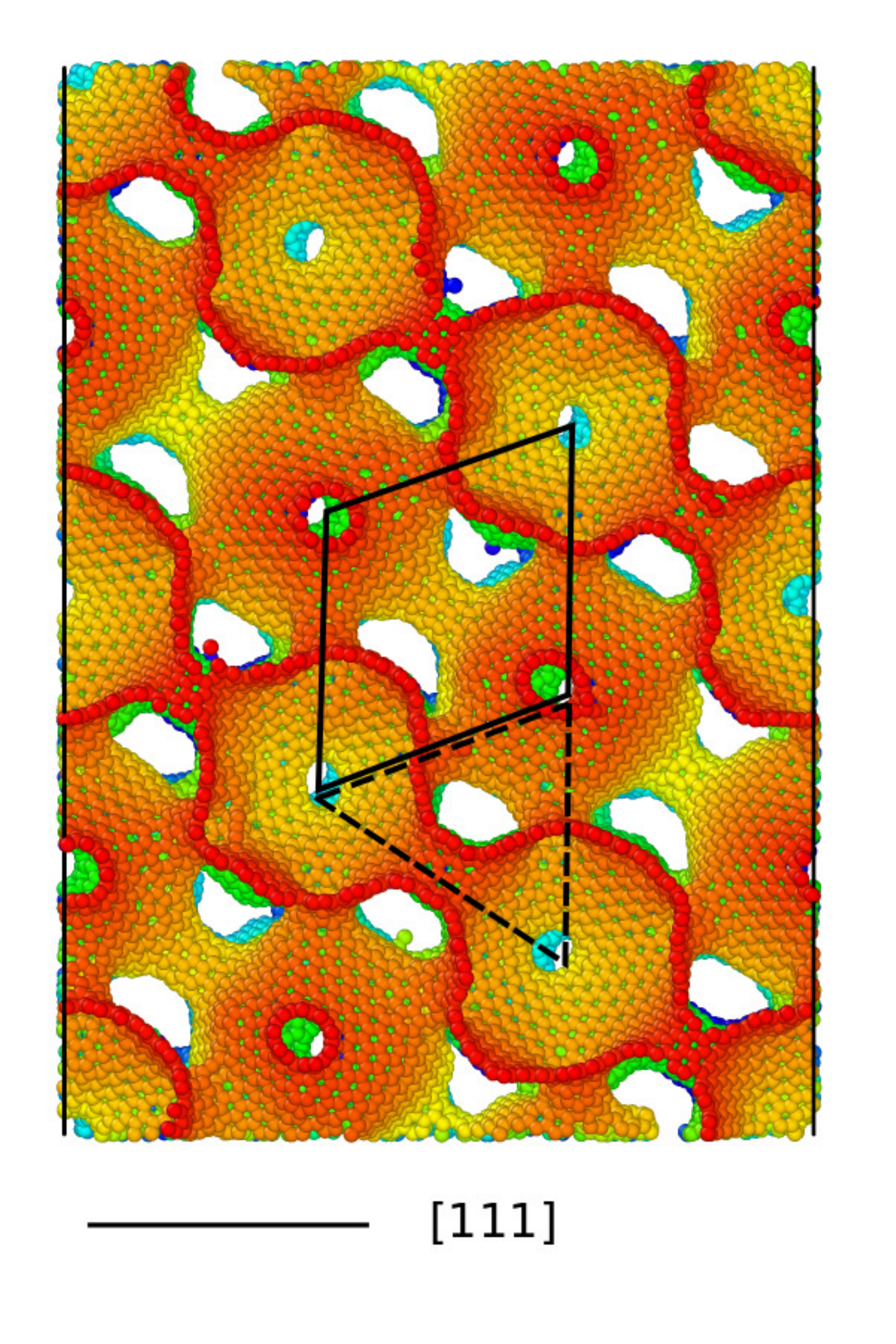}}
\quad
\subfigure[Strain 20\%]{
\includegraphics[width=0.30\textwidth]{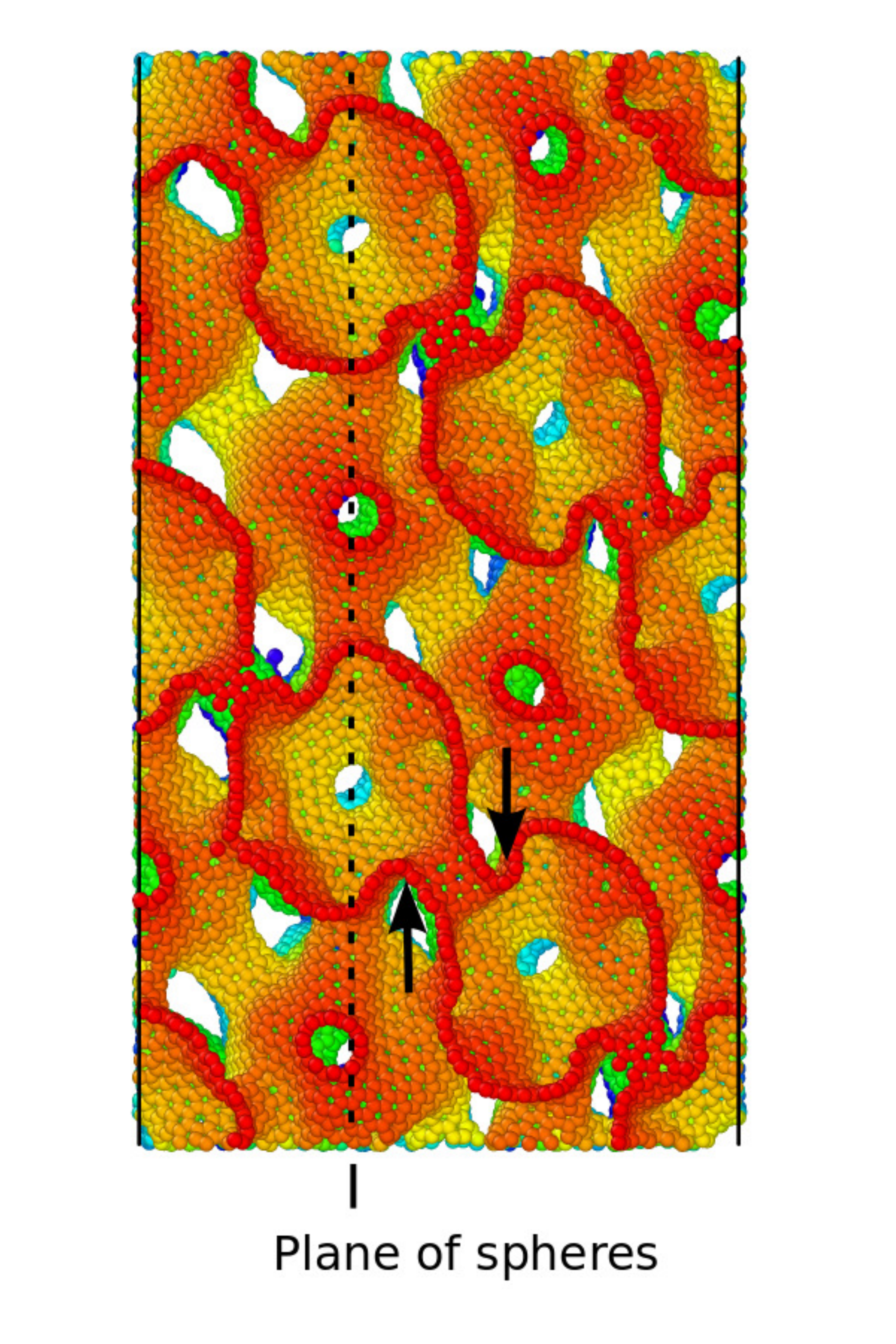}}
\caption{Nanotruss network number 3 under uni-axial compression along the [111] direction. The first image is the undeformed sample; the second is a snapshot at $20$\% strain. The centers of the graphene spheres divide the space in two volumes of different shape: an octahedron (rendered in red in Fig. \ref{fig:1}b) and a tetrahedron (rendered in blue in Fig. \ref{fig:1}b). These two volumes are reported in image a). It is clear from image b) that the tetrahedrons are loaded from one vertex towards the center of their base so that the nanotubes are forced to rotate during the compression. Notice the collapse of the planes of spheres in the second image. In particular, for each plane collapsed there is a peak in the stress-strain curve as reported in Fig. \ref{fig:8}. Colors have been used for visualization purposes only and have no physical meaning.}.
\label{fig:9}
\end{figure*}

The four samples reported in Fig.~\ref{fig:8} behave similarly in the elastic regime. However, samples 2 and 3 present an inverted response with respect to traction, and the stress for sample 2 in compression is higher than that for sample 3.
In the plateau, a geometry-related response occurs. Sample 1 shows a nearly flat plateau that does not approach the densification regime up to $50$\% strain. Samples 2 and 4 present a collapsing plateau before approaching the densification regime at high strain. For samples 1, 2 and 4 one can see from the unloading part of the stress-strain curve that the deformation is partially plastic, without full recovery of the original shape at zero stress. This is a clear signature of mechanical hysteresis in these systems.

Sample 3 presents a response that is very different from the other samples, showing a number of peaks in the plateau of the stress-strain curve (Fig.~\ref{fig:8}). Each peak is related to the collapse of a plane, as shown in Fig.~\ref{fig:9}.
This type of response, peculiar to sample 3, has been found only along the [111] direction. It is worthwhile noting that this response depends on the ratio between the nanotube's and the sphere's diameters. The collapse of the planes is essentially elastic, indeed the sample presents only a small residual deformation at zero stress.

\begin{figure}[] 
\centering
\includegraphics[width=0.6\textwidth]{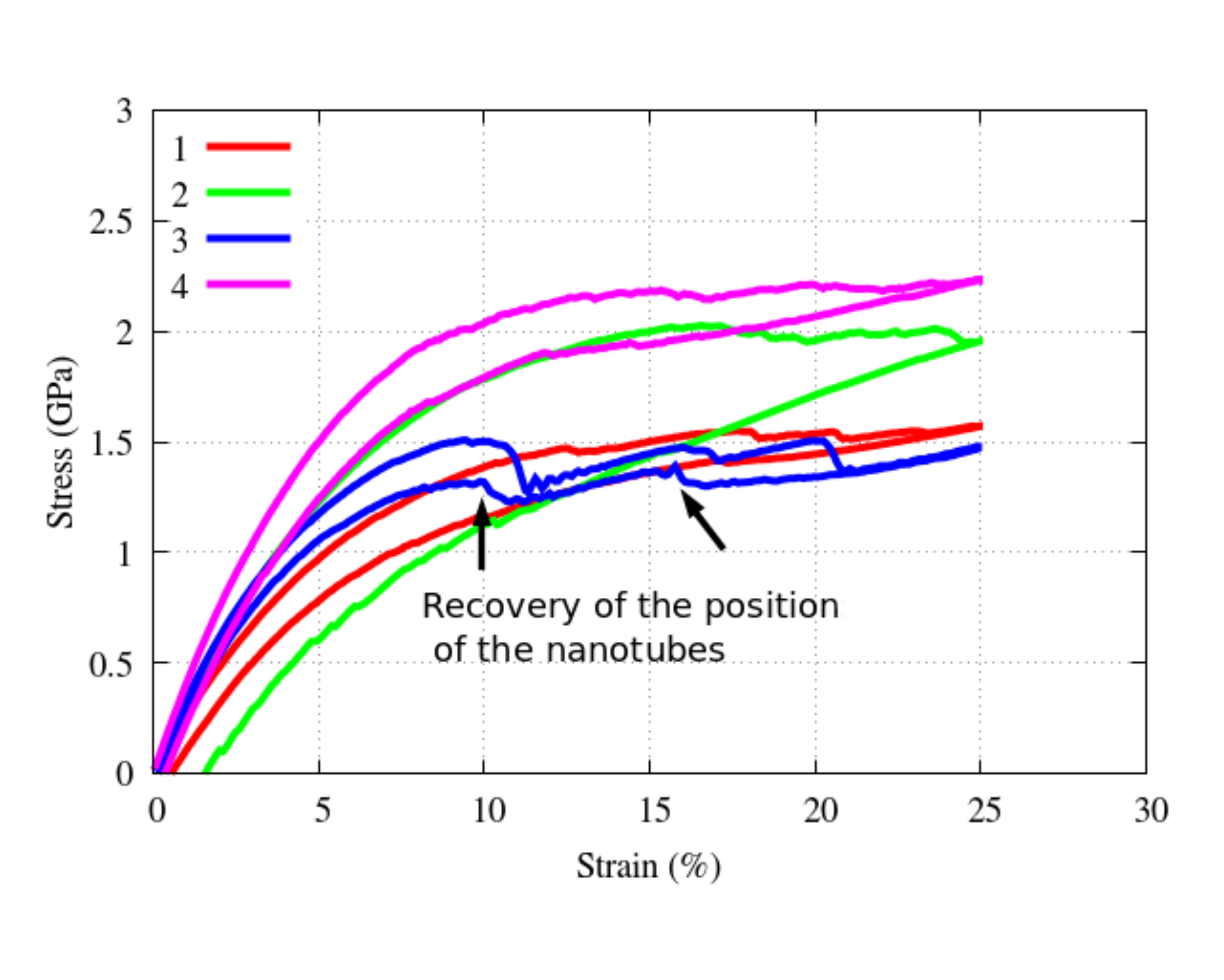}
\caption{ Stress-strain curves of the four nanotruss networks under uni-axial compression along the [111] direction up to $25$\% strain. Sample 3 presents a stress-strain curve qualitatively different from the other samples. It presents three peak in the loading part of the curve and two peaks in the unloading part. These peaks are due to the global instability of the planes of spheres, as shown in Fig. \ref{fig:9}.}
\label{fig:10}
\end{figure}

Finally, we note that the partial insertion inside the sphere of the nanotubes that carry the compressive load is caused by a global instability of the sphere's planes: all the nanotubes that are loaded in compression are influenced in the same manner. 

The centers of the graphene spheres divide the space in two volumes of different shape: an octahedron (rendered in red in Fig. \ref{fig:1}b) and a tetrahedron (rendered in blue in Fig. \ref{fig:1}b). These two volumes are reported in panel a) in Fig. \ref{fig:9}. 
The tetrahedrons are loaded from one vertex towards the center of their base, forcing the nanotubes to rotate during the compression. The load acts in the direction of stabilizing the structure as, remarkably, the samples do not present reciprocal sliding of spheres' planes. This means that further simulations in this load direction could be performed on a single unit cell without loss of generality. Indeed, for samples in which is present a transverse sliding of the planes the boundary conditions as well as the parity of the number of planes in the simulation cell play a critical role. In the simplest case of a unit cell with only a plane, for example, the sliding is completely prevented. 

In Fig.~\ref{fig:10} the compressive stress-strain curves up to 25\% strain are reported. The unloading part of the curve for sample 3 presents some stress peaks due to the repositioning of nanotubes into their original positions. The number of oscillations is related to the number of planes in the sample; therefore, in the limiting case of a bulk material, we expect that the number of oscillations will increase reaching a flat plateau.

Full recovery after compressive deformation similar to that found for sample 3 has been reported, for example, in carbon nanotube bundles \citep{Cao2005} and in ultralight boron nitride foams \citep{Yin2013}; in these systems, the deformation in absence of plastic strain is allowed by rotation of the internal components. 
 
In Tab. \ref{tab:Table Comp} we report the Young modulus, plateau stress and residual deformation of the four samples, under uni-axial compression in the [111] direction. 

In the next section we will focus on a quantitative evaluation of the Poisson ratio for the samples 3 and 4 in the [110] direction, for which an auxetic response emerges.

\subsection{Negative Poisson ratio}

\begin{figure*}[]  
\centering

\subfigure[Strain 0\%]{\includegraphics[width=0.25\textwidth]{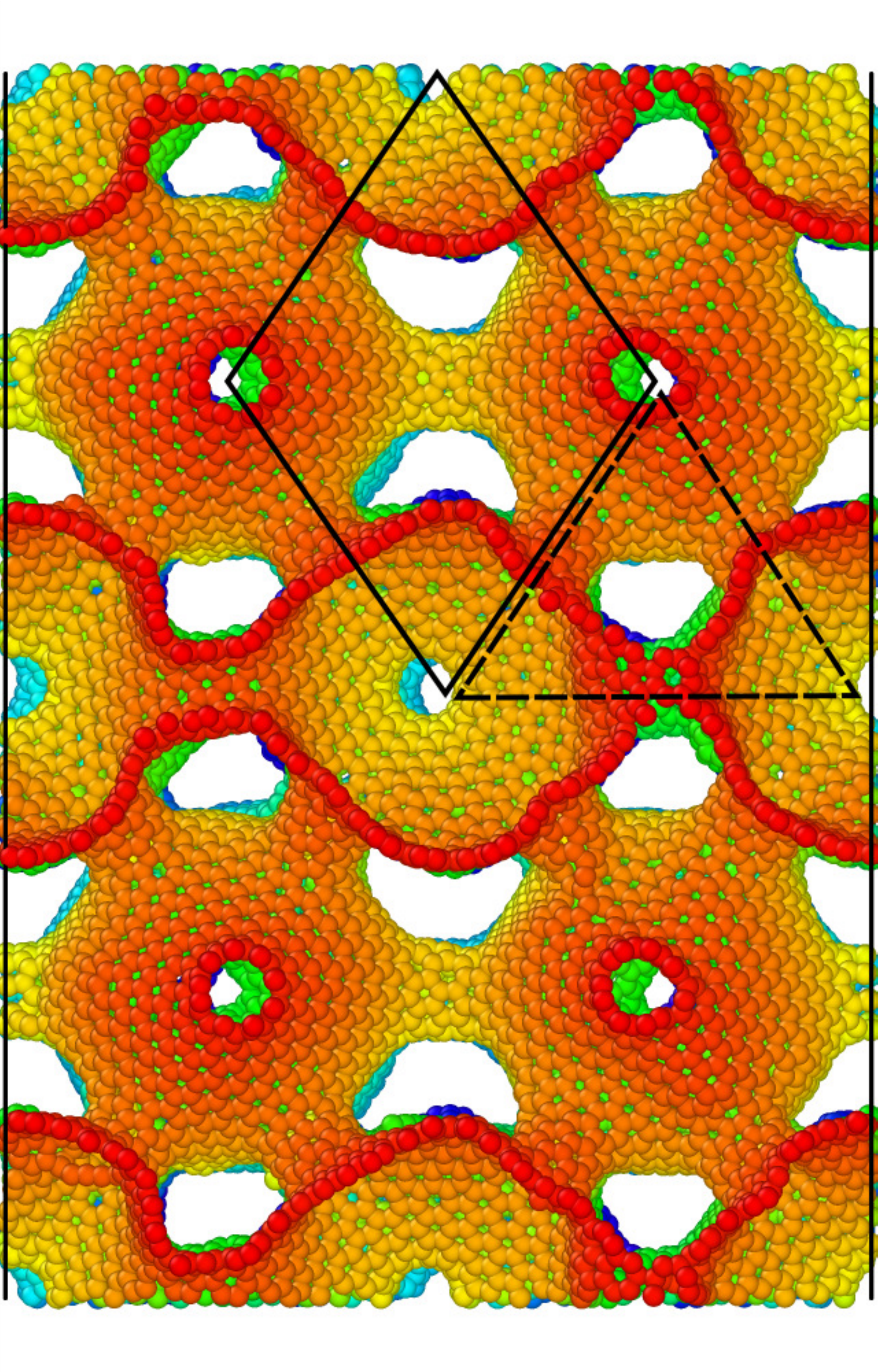}}
\quad
\quad
\subfigure[Strain 20\%]{\includegraphics[width=0.25\textwidth]{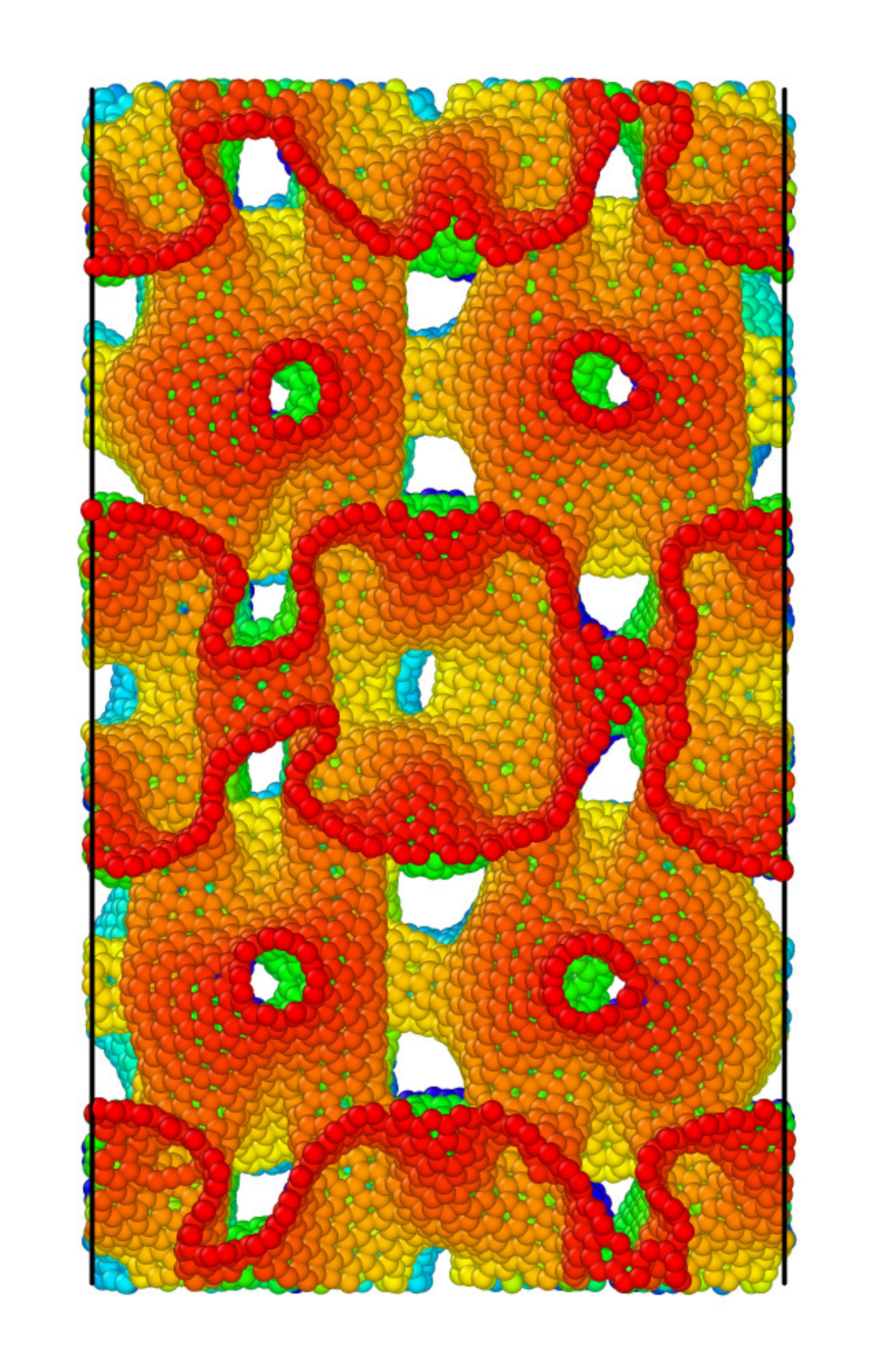}}
\subfigure[Strain 30\%]{\includegraphics[width=0.25\textwidth]{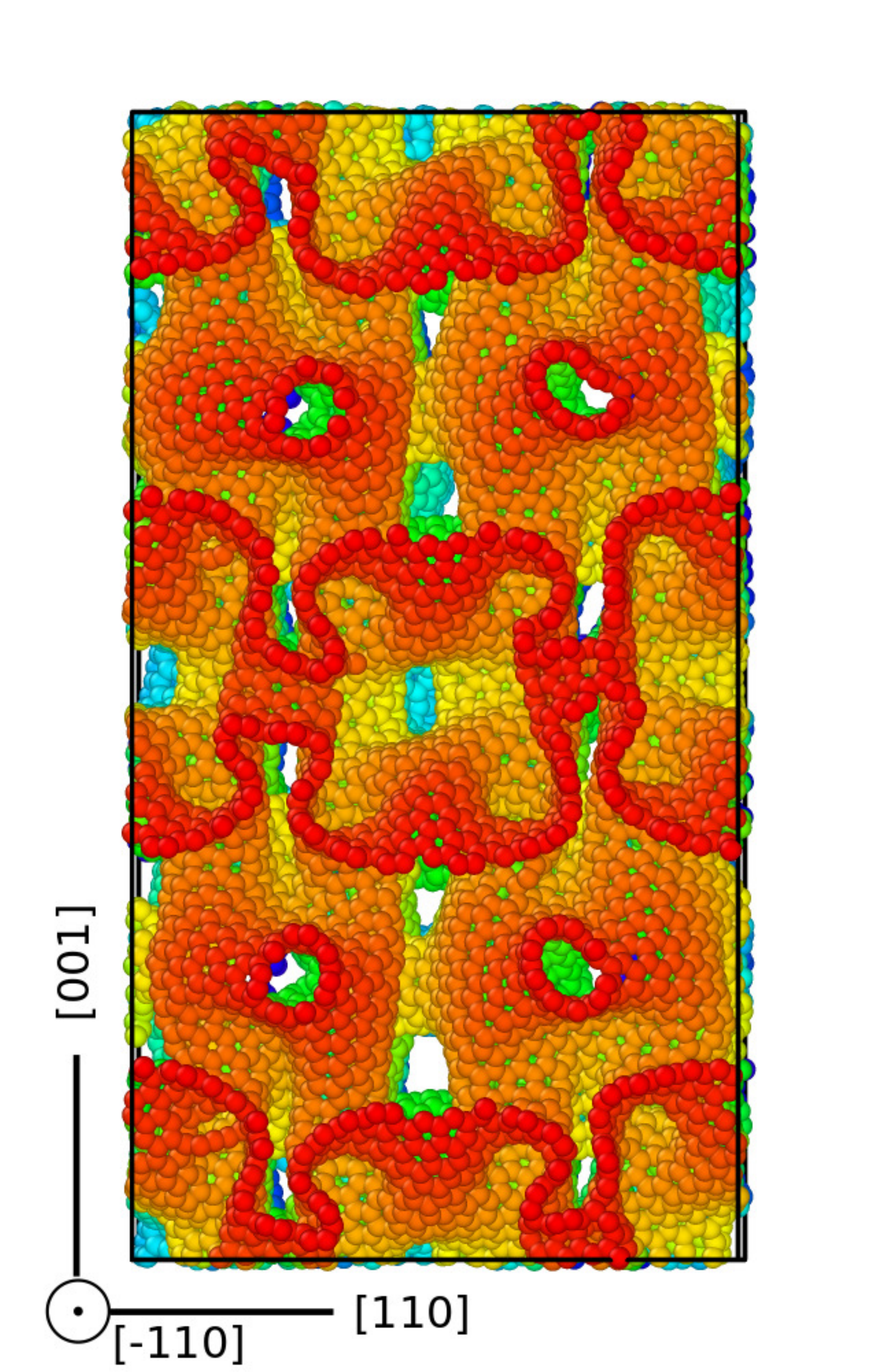}}

\caption{Three snapshots of sample 3 under uni-axial compression in the [110] direction at $0$~\%, $20$~\% and $30$~\% strain. In image a) are shown the positions of an octahedron and a tetrahedron. In this case the load is aligned with an edge of each tetrahedrons. Due to the nanotubes insertion there is a contraction in the transverse direction [001] direction. The Poisson ratio with respect to this direction has negative values for strain higher than $10$~\%. With respect to the [-110] direction (out of the page) the Poisson ratio is initially positive and then tends to zero as strain increases to $30$~\%. The images are to scale.}
\label{fig:11}
\end{figure*}

\begin{figure*}[]  
\centering
\subfigure[Strain 0\%]{\includegraphics[width=0.40\textwidth]{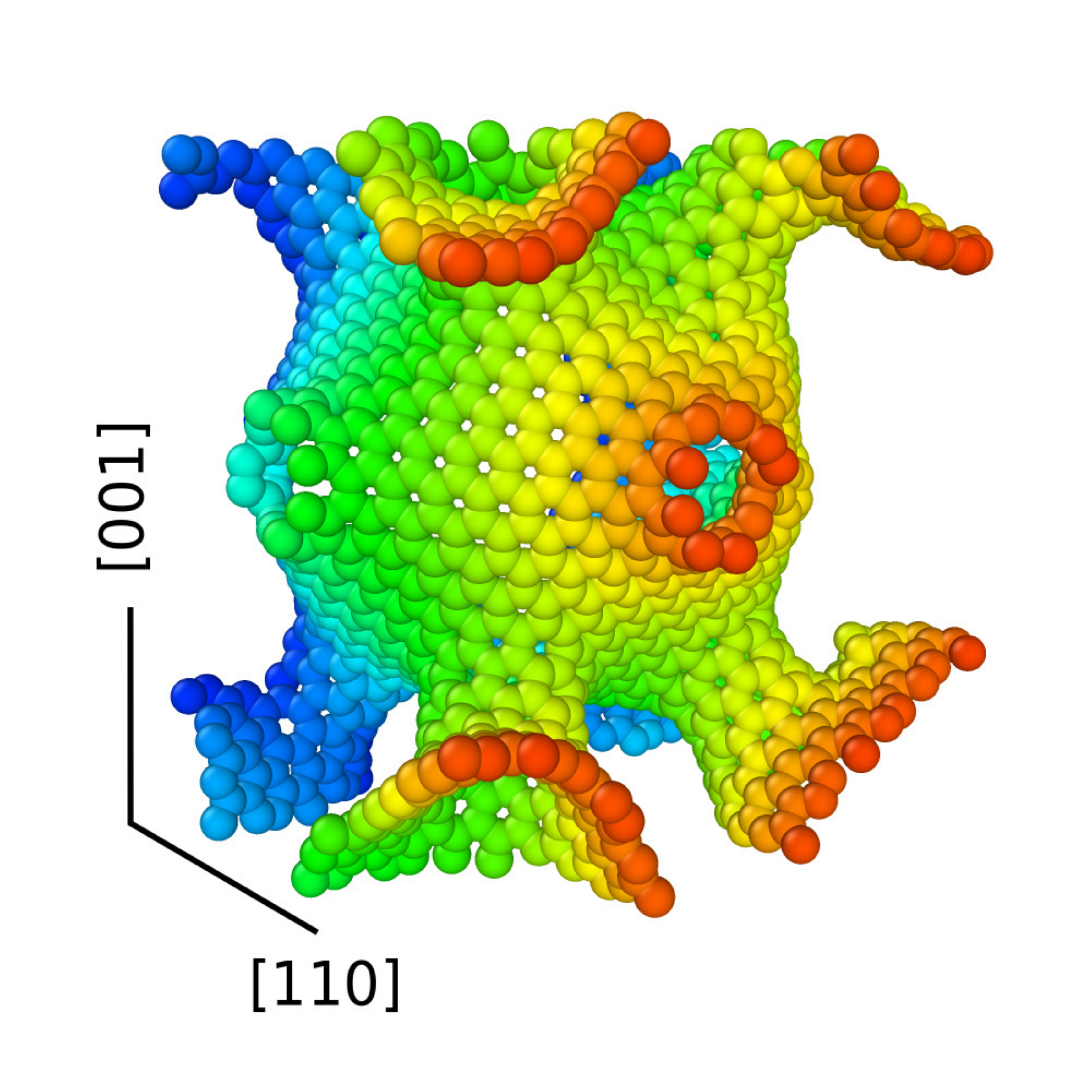}}
\subfigure[Strain 20\%]{\includegraphics[width=0.40\textwidth]{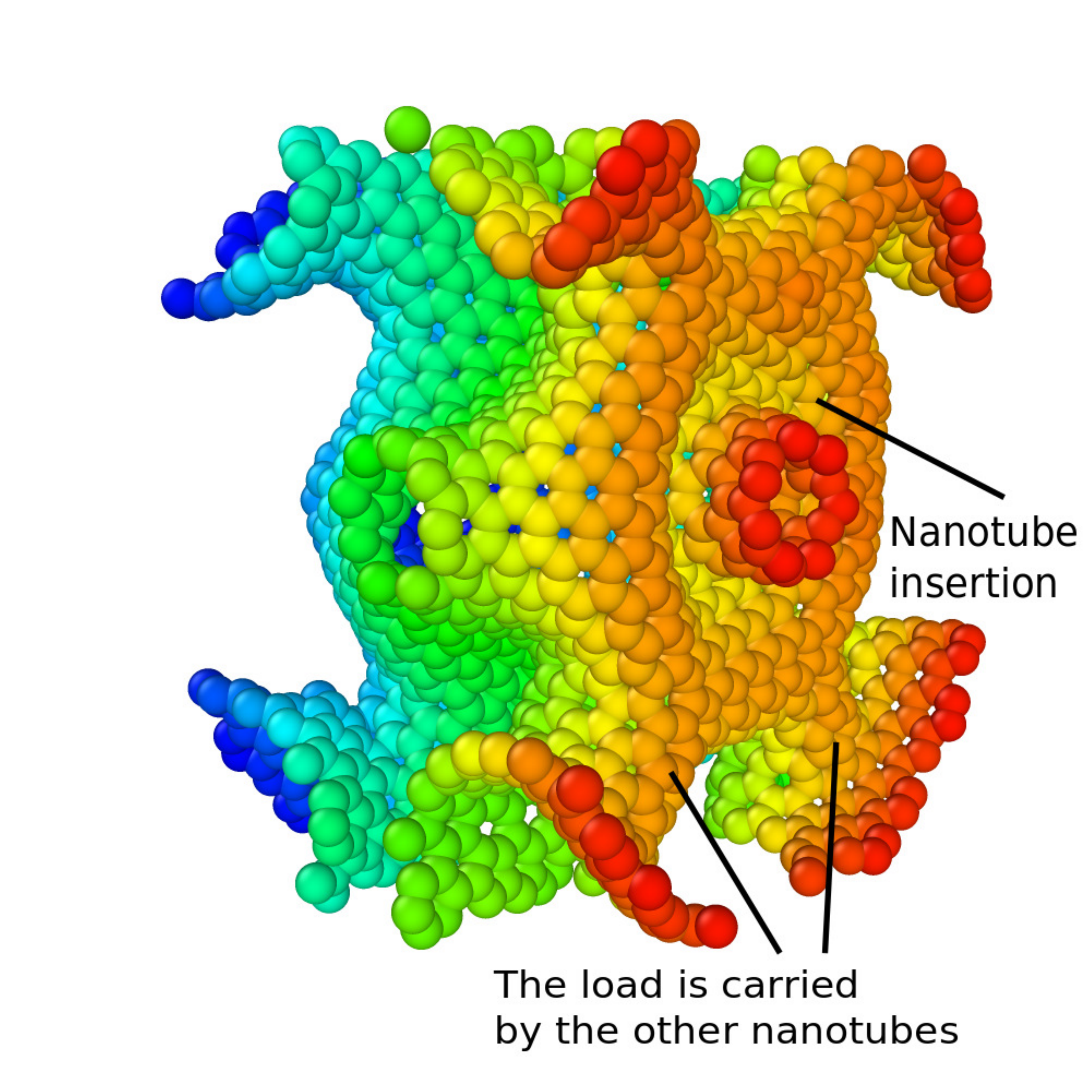}}
\caption{The two snapshots show a single sphere of sample 3 under uni-axial compression in the [110] direction at $0$~\% and $20$~\% strain. Panel a) shows the undeformed sphere. Panel b) reports the sphere during the deformation.}
\label{fig:12}
\end{figure*}

The quantity characterizing the materials response in the direction orthogonal to compression is the Poisson ratio $\nu$, defined as the ratio between the negative transverse and longitudinal strain.
Commonly, materials have a positive Poisson ratio, meaning that they expand in the direction orthogonal to the external compressive load (e.g. steel, $\nu=0.3$).
Under compression in [110] direction we found negative Poisson ratio for samples 3 and 4. 

This negative Poisson ratio is strictly connected to an interesting feature, not present in the compression along the other directions, that can be found in the [110] direction. This characteristic, sketched in Fig.~\ref{fig:11} for sample 3, is the longitudinal insertion of the nanotubes inside the spheres. Nevertheless, the stress-strain curve in the [110] direction does not show particular engaging features, as the load is carried by the other nanotubes. This local instability, related to a single nanotube and not to a whole plane of spheres, has been also found in sample 4 along the [110] direction. 

Our simulations also show that the nanotube insertion presents three stable configurations. In two of these configurations, only the bottom part of the nanotube enters one of the adjacent spheres, while in the third one both top and bottom sides of the nanotubes are inserted. We note that this behavior introduces a new degree of freedom under compression for each nanotube aligned with the [110] direction.

In Fig.~\ref{fig:12} we report two snapshots of a single sphere 
of the sample 3 under compression. The insertion increases the local curvature of the surface, and, as for the [111] direction, this could be useful to enhance the adsorption or desorption of gases. Due to the symmetry of the structure, the nanotubes in the horizontal plane perpendicular to the [110] direction are free of moving under a second load applied in the direction [-110]. For longer nanotubes, the inserted extremities could interact each other giving collective patterns in the horizontal plane, in particular under bi-axial compression (in directions equivalently to the [110] and [-110] directions).

\begin{figure}[] 
\centering
\includegraphics[width=0.6\textwidth]{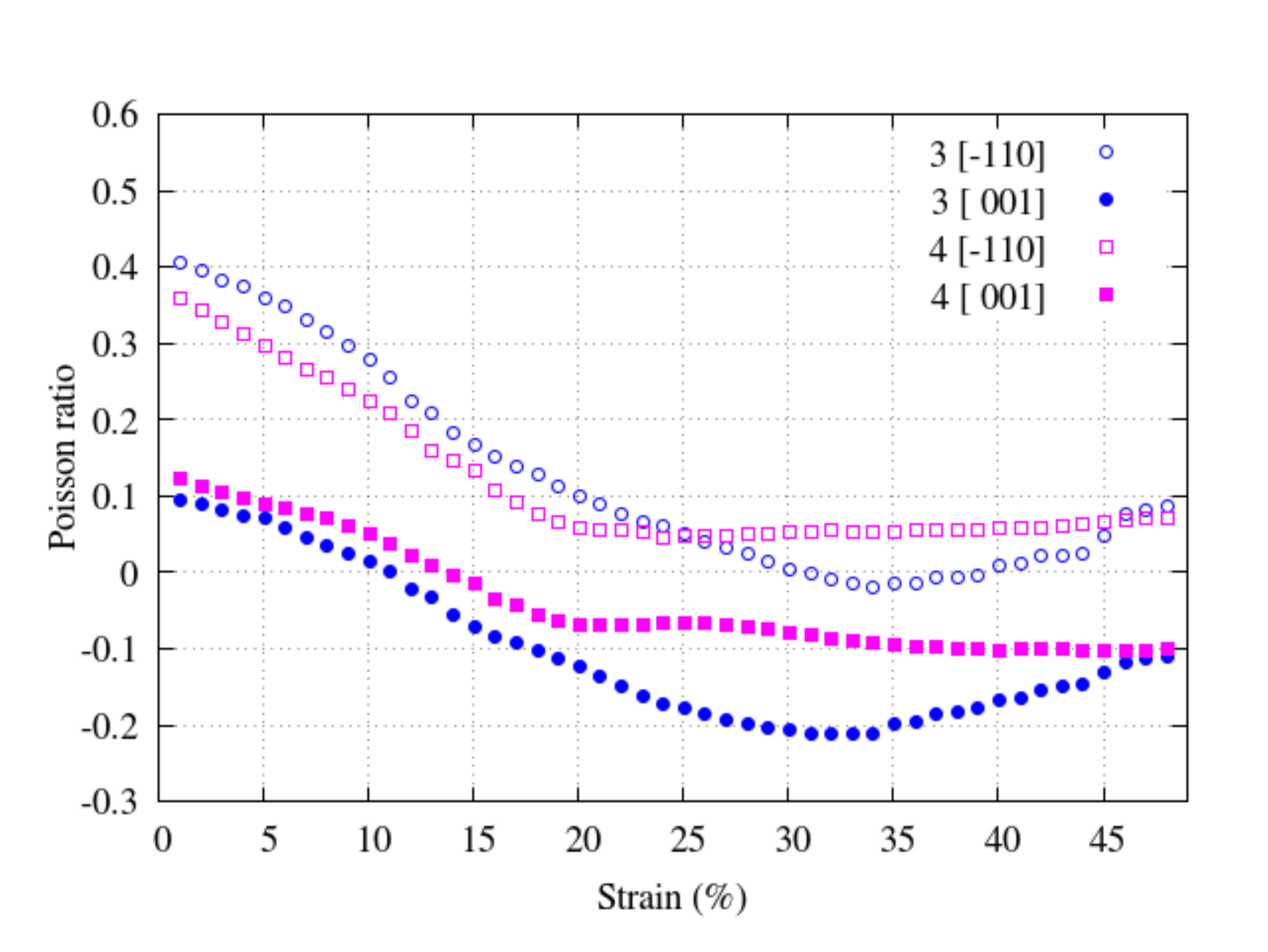}
\caption{Poisson ratio of samples 3 and 4 under compression in the [110] direction. The Poisson ratios are between the [110] direction (of load) and the other two directions, [-110] and [001] as indicated by the labels. See Fig.~\ref{fig:11} for graphical representation of the axes. These two samples present an auxetic response (negative Poisson ratio) for large deformations.}
\label{fig:13}
\end{figure} 

In Fig.~\ref{fig:13} we report the Poisson ratio for samples 3 and 4 under uni-axial compression in the [110] direction. The Poisson ratio is monotonically decreasing in both the samples. Indeed, these two structures are characterized by positive values of $\nu$ for small strain in the [-110] and [001] directions. At higher strain, the Poisson ratio of the sample 3 along the [-110] and [001] directions presents a minimum at around $35$~\% strain. A similar trend is found for the sample 4 with the presence of a plateau starting from $20$~\% strain. Furthermore, we note that Poisson ratios along the z direction become negative. A negative Poisson ratio means that, if a compressive load is applied in the [110] direction, the material tends to increase its density under the applied load. This feature of carbon foams could be useful for delivering effective high-impact energy absorption. Finally, we note that in this case the compressive load tends to stabilize the structure only for small strain, at variance, for high strain, a sliding of the spheres' plane is found.

\subsection{Density scaling relation}

The Young modulus $E$ of microstructures obeys a semi-empirical power-scaling law with relative density as follows \citep{Ashby2006}:
\begin{equation}
\frac{E}{\tilde{E}} = \alpha 	\left(\frac{\rho}{\tilde{\rho}}\right)^\beta \label{Scaling}
\end{equation} 
where $\tilde{E}$ and $\tilde{\rho}$ are the Young modulus and the density of the bulk material respectively, $\alpha$ is a proportionality coefficient and $\beta$ depends from the micro-architecture of the material. In our case $\tilde{E}$ is the Young modulus of graphene, which composes the nanotrusses.
FCC nanotrusses, where each node is connected to $12$ first neighbours with similar struts, present stretching-dominated response \citep{Deshpande2001}. Indeed, the octet-truss satisfies Maxwell's criterion for static determinacy. This criterion in three dimensions is given by 
\begin{equation}
b-3j+6 \geq 0
\end{equation} 
where $b$ and $j$ are the number of struts and nodes, respectively, in the unit cell. For this type of octet-truss lattice material \citep{Deshpande2001} $\alpha=0.3$ and $\beta$ has value $1$. Differently, for bending-dominated structures such as foams, the scaling relation Eq. \ref{Scaling} is satisfied with $\alpha=1$ and $\beta=2$. 
We plan to use the scaling relation Eq. \ref{Scaling} to estimate the average Young modulus $\tilde{E}$ of the graphene sheets that compose the nanotruss networks, using as coefficients and exponents those valid in the limiting cases of linear and parabolic scaling laws. 
Furthermore, we use the bulk density $\tilde{\rho}$ of graphite ($\tilde{\rho}=2250$~kg~m${^{-3}}$). Due to the presence of defects in the nanotrusses one can expect a reduction of the Young modulus of our structures with respect to graphene ($\tilde{E}=1.0$~TPa  \citep{Zhao2009}) to values typically found in nanometric graphene sheets (in the range $100-500$~GPa \cite{Becton2015}).    

\begin{table}[] 
\centering
\small
\begin{tabular}{cccccc}
\toprule

       &Relative  & Young modulus   & Young modulus   \\ 
Sample &density   & linear relation  & parabolic relation \\ 
       &    (\%)        &    (GPa)  &            (GPa)               \\
\midrule
1 & 28.2 & 319.4 &     340.1 \\ 
2 & 31.4 & 350.1 &     334.2 \\ 
3 & 31.3 & 383.0 &     366.7 \\ 
4 & 30.8 & 455.2 &     444.0 \\ 
\bottomrule
\end{tabular}
\caption{Relative density with respect to graphene, Young modulus of graphene that constitutes the nanotruss networks calculated by means of the relation in Eq. \eqref{Scaling} with $\alpha=0.3$ and $\beta=1$ (linear relation) or $\alpha=1$ and $\beta=2$ (parabolic relation).}
\label{tab:Table Coeff}
\end{table}

In Tab. \ref{tab:Table Coeff} we report the computed Young modulus of graphene in the case of linear and parabolic density scaling law. We note that the estimated Young modulus is indeed lower than pristine graphene and is in the range $100-500$~GPa.

\subsection{Energy absorption}

Due to the possible application of nanotruss networks to impact absorption, it is important to study the mechanical behavior of nanotruss and compare that behavior with standard materials for energy absorption.

The compressive stress-strain curves can be used to assess the performance of a material with regard to its energy absorption properties. Owing to the similarity of the stress-strain curves of our samples to those typical of absorber materials, we evaluated the most relevant quantities used to describe this property, such as:\\
the \emph{crush force efficiency}
\begin{equation}
\eta(\varepsilon) = \frac{F_\mathrm{av}(\varepsilon)}{F_\mathrm{max}(\varepsilon)}
\end{equation} 
where $F_\mathrm{av}(\varepsilon)$ is the average of the stress between $0$ and $\varepsilon$ strain and $F_\mathrm{max}(\varepsilon)$ is the maximum stress up to strain $\varepsilon$. For the ideal energy absorber $\eta=1$.
\\
The \emph{stroke efficiency}
\begin{equation}
S_{E} = \varepsilon_\mathrm{dens}
\end{equation} 
that represents the strain on the edge of densification and gives an estimation of the ratio between the useful length to absorption and the total length. \\
The \emph{specific energy absorption} is the energy absorbed per unit of mass 
\begin{equation}
SEA = \frac{E_\mathrm{t}}{m}
\end{equation} 
where $m$ is the mass and the \emph{total energy} absorbed
\begin{equation}
E_t(\varepsilon)=V \int_0^{\varepsilon} \! \sigma(\varepsilon') \, \mathrm{d}\varepsilon'
\end{equation} 
is the area under the stress-strain curve up to strain $\varepsilon$ times the sample's volume $V$. 

All previous quantities, except the stroke efficiency, are functions of the strain and in order to compare different materials they are usually evaluated at the densification strain $\varepsilon_\mathrm{dens}$.
The densification strain is evaluated so to maximize the \emph{energy absorption efficiency} \citep{Li2006, Avalle2001}

\begin{equation}
\eta_t(\varepsilon)= \frac{1}{\sigma(\varepsilon)}\int_0^{\varepsilon} \! \sigma(\varepsilon') \, \mathrm{d}\varepsilon'
\end{equation} 

\begin{table*}[] 
\centering
\small
\begin{tabular}{cccccc} 
\toprule
             & Crush force &     Stroke    & Specific energy           & Specific energy          &   \\ 
 Sample      &  efficiency &   efficiency  & absorption compression    & loading-unloading        &  \\ 
             &     $\eta$        &    $S_{E}$   & 50\%  (MJ~kg$^{-1}$)      & 50\%  (MJ~kg$^{-1}$)     &  \\
\midrule
1 &   0.7 & 47 &  1.2 & 0.5 \\ 
2 &   0.6 & 43 &  1.4 & 1.0 \\ 
3 &   0.5 & 42 &  1.2 & 0.3 \\ 
4 &   0.7 & 40 &  1.6 & 0.8 \\ 
\bottomrule
\end{tabular}
\caption{Crush force efficiency, stroke efficiency, and specific energy absorption calculated up to 50\% strain for the four samples under uni-axial compression in the [111] direction.}
\label{tab:Table Absorption}
\end{table*}

These quantities are reported in Tab.~\ref{tab:Table Absorption} for the four samples compressed in the [111] direction. The amount of absorbed energy is reported at 50\% strain that is a standard strain for the evaluation of the performance of energy absorption materials. This value is higher than $\varepsilon_\mathrm{dens}$ for some samples, however for comparative purposes is preferable to choose a common value for strain. The best performances are obtained for sample 4. We compare these values to those of typical absorbers, such as aluminum foams, high strength steel or bulk aluminum. The order of magnitude of the specific energy absorption in aluminum foams, high strength steel and bulk aluminum is of the order of $10^{-2}$~MJ~kg$^{-1}$, thus much lower than that of the carbon nanotruss networks we studied in this work.

\section{Conclusions}

The mechanical properties of carbon nanotruss networks in both tensile and compressive regimes were investigated from atomistic simulations by using a reactive potential. Our simulations were performed by stretching various nanotruss configurations along the symmetric orientations [100], [110], [111].

First of all we discussed a method suitable for obtaining realistic structures taking into account the presence of defects. 
We used the stiffness matrix as a tool to analyze the anisotropy of the samples, finding a small anisotropy due to the nano-structure of our material and the limited box length of the unit cell used during the Voronoi dualization. 
The presence of pre-strained graphene sheets inside the sample can change the mechanical properties as the Young modulus, when this is computed for small deformations. We find that this is principally due to the pre-strained nature of our realistic configurations. 

Mechanical properties of nanotruss networks are strongly influenced by the presence of defects because these are the points at which cracks nucleate. Nanotruss networks presents a brittle nature inherited from the parent material, that is graphene. Furthermore, we found a parabolic trend of the stress-strain curves under tensile load and a tensile strength is roughly proportional to the Young modulus, as well as a Young modulus comparable to that of bulk metals like aluminum.

Our results show the emergence of different responses in compression for different orientations and different truss geometry. In particular, we have found that the non linear response related to the partial insertion of the nanotubes inside the spheres is screened in the [111] direction. In this case, a suitable choice of the lattice parameters (as done in sample 3) allows a plateau to be obtained in elastic regime. This elastic response is amenable to develop materials that allow the absorption of energy without plastic deformation. 

A local unstable insertion of the nanotubes in the spheres was found in the [110] direction for a proper choice of the truss geometry. The first type instability influences the mechanical properties, whereas the main effect of the second instability is to deliver a negative Poisson ratio, as found in particular configuration of nanotrusses (samples 3 and 4). 
Density scaling relations for nanotruss networks were analyzed assuming a linear or a parabolic relation between relative density and relative Young modulus to estimate the Young modulus of defective graphene that constitutes the nanotruss networks. A range of possible values in accordance with the literature was found.

Finally, energy absorption properties of these carbon nanotrusses have been studied. We found that nanotruss networks could indeed outperform typical materials used as means of energy absorption, such as aluminum foams, high strength steel or bulk aluminum due to an increased specific energy absorption.

Based on the results of this study, nanotruss networks may represent candidates for porous, flexible, and high strength materials. Fields as impact energy absorption as well as structural mechanics can take advantage from the novel properties shown by carbon nanotruss networks.

\section*{Acknowledgments}
S.T and G.G. acknowledge funding from European Commission under the Graphene
Flagship (WP10 “Nanocomposites”, no. 604391). A.P. and S.T.
acknowledge Dr. R. Gabbrielli and Prof. A. Iorio for useful
discussions on Voronoi patterning of Lennard-Jones 
optimized networks. A.P. acknowledges Dr. M. Christian 
for reading this manuscript. NMP is supported by the European
Research Council (ERC StG Ideas 2011 BIHSNAM no.
279985 on “Bio-Inspired hierarchical super-nanomaterials”,
ERC PoC 2015 SILKENE no. 693670 on “Bionic silk with
graphene or other nanomaterials spun by silkworms”, ERC
PoC 2013-2 KNOTOUGH no. 632277 on “Super-tough knotted
 fibres”), by the European Commission under the Graphene
Flagship (WP10 “Nanocomposites”, no. 604391).
\\

\section*{References}
\bibliography{Article}

\end{document}